\documentclass[preprintnumbers,article,amsmath,amssymb,floatfix,10pt,prd,onecolumn,
superscriptaddress,nofootinbib]{revtex4-2}
\usepackage{amsmath}
\usepackage{eurosym}
\usepackage{epstopdf}
\usepackage{capt-of}
\usepackage{graphicx}
\usepackage{dcolumn}
\usepackage[left=0.8in, right=0.8in, top=1in, bottom=1in]{geometry} 
\RequirePackage{graphicx}
\RequirePackage{float}
\RequirePackage{hyperref}
\RequirePackage{amsmath}
\RequirePackage{amssymb}
\RequirePackage{mathtools}
\usepackage{color}
\usepackage{comment}
\usepackage{xcolor}

\begin{document}
\color{black}       
\title{Observational constraints using Bayesian statistics and Deep Learning in $f(Q)$ gravity}
\author{Lokesh Kumar Sharma}
\email{lokesh.sharma@gla.ac.in}
\affiliation{Department of Physics, GLA University, Mathura 281406, India}

\author{Suresh Parekh}
\email{thsureshpareh@gmail.com}
\affiliation{Department of  Physics, SP Pune University, Pune 411007, Maharastra, India}

\author{Anil Kumar Yadav
}
\email{abanilyadav@yahoo.co.in}
\affiliation{Department of Physics, United College of Engineering and Research, Greater Noida - 201310, India}

\begin{abstract}
\noindent \textbf{Abstract:} This study investigates the evolution of Friedmann-Robertson-Walker (FRW) cosmological models within the $f(Q)$ gravity framework, utilizing a specific $f(Q)$ formulation and a novel Hubble parameter $H(z)$ parameterization to probe the universe's accelerating expansion.
A central aspect is the application of advanced machine learning techniques for cosmological parameter estimation, alongside comparisons with traditional Bayesian (MCMC) methods.
We employ a hybrid Mixed Neural Network (MNN), which synergistically combines Artificial Neural Networks (ANNs) and Mixture Density Networks (MDNs), to enhance the accuracy and robustness of parameter constraints.
This MNN architecture is integrated into the CoLFI (Cosmological Likelihood-Free Inference) framework.
CoLFI facilitates likelihood-free inference, a significant methodological advancement that provides an efficient and robust alternative, particularly for complex models with computationally expensive or intractable likelihood functions.
Training efficiency for the neural networks is optimized by generating data via hyperellipsoid sampling.
The $f(Q)$ model, constrained using these diverse approaches, successfully describes a universe transitioning from an early decelerating phase to the current accelerated expansion, with a computed transition redshift of $z_t = 0.60$.
The physical and kinematic properties of the model are discussed, underscoring the efficacy of the MNN-CoLFI methodology and its consistency with MCMC results, while highlighting its advantages for obtaining observational constraints in $f(Q)$ gravity.
\end{abstract}
\maketitle
\section{introduction}
A crucial idea for understanding certain hidden aspects of gravitational components that provide a basic understanding of astronomical wonders and the cosmos is Einstein's general relativity. It is certain that the cosmos experiences an early expansion, and late-time accelerated expansion has recently been confirmed by a few observable findings \cite{1,2,3}. Some have hypothesized that this is caused by dark energy (DE), an enigmatic force with catastrophic gravitational effects. The peculiarities of DE and the problems with astronomical acceleration may be handled primarily in two ways. Alternative approaches to dealing with fuse DE include different gravity assumptions, which lead to a change in the activity standard in general relativity. Several dynamical DE candidates are offered in the alternative approach to help understand DE \cite{4, 5}. Introducing a cosmic-consistent identifier with the state of state limit $\omega = -1$ is the most straightforward way to understand this DE. Moreover, in the summarized literature on cosmological dependable DE, there are other contenders, including holographic and age-realistic DE, which are linked to the energy thickness of dynamical scalar fields \cite{6, 7,8,9,10,11,12,13,14,15,16}. However, it is worth to noting that the limitations occurs in General Theory of Relativity (GTR) at various energy scales question its validity for being the best theory that depicts the gravitational interactions. Indeed, at the quantum scale, the GTR exhibits trouble with its extension to a quantum theory of gravity, since GTR is non re-normalizable, and forecasting of infinite gravitational tidal forces with existence of different singularities. Therefore, in GTR framework, one can not described the late time accelerating expansion phenomenon of the universe within the context of the standard model of cosmology without introducing the cosmological constant $\Lambda$, whose theoretical value evaluated by quantum field theory differs from the value inferred by the Friedman equations. Thus, one an argue that the cosmological constant $\Lambda$ in form of dark energy is driving the late time acceleration of the universe but the existence of dark energy, to date, is only based on indirect observations. Furthermore, it is worthwhile to mention that the dynamics of the farthest stars orbiting around the center of galaxies represents an other open issue currently addressed to a never-detected form of matter called dark matter. The incompatibilities mentioned above redirect beyond GTR, and therefore open up the possibility of modified gravity. Some important applications of modified theory of gravity in various physical contexts are given in Refs. \cite{Nashed/2023,Nashed/2022,Yousaf/2024a,Malik/2024,Heisenberg/2023,Yousaf/2024b}.  

Furthermore, the $G$, $R$, $Q$ and $T$ are the symbols that are used in modified theories of gravity to denote the Gauss-Bonnet invariant, Ricci, non-metricity scalars of the universe, and trace of the energy-momentum tensor. Various exercises have been created, and intriguing discoveries have been made within the framework of these altered theories of gravity \cite{17,18,19}. The authors of Refs. \cite{Jimenez/2018,Beltran/2019} have investigated a novel form of $f(Q)$ theory that attributes gravity to non-metricity and operates within the framework of Weyl geometry. Moreover, we also note that, like $f(T)$ and $f(R)$ theory of gravity, the $f(Q)$ gravity also exhibits  a generalization of the Teleparallel Equivalent of General Relativity. Therefore, in recent times, the $f(Q)$ theory of gravity is used to explore to describe the various aspects and fate of the universe. For example, Albuquerque et al. \cite{Albuquerque/2022} have examined the  linear perturbations and its  implications in $f(Q)$ theory of gravity. A further point to consider is that Bajardi et al. \cite{20} used an order reduction strategy in order to limit the functional form of $f(Q)$ in extended symmetric teleparallel cosmology. In a non-linear fashion, the non-metricity scalar for MOND is extended in reference \cite{21}. The EoS $(\omega)$ in $f(Q)$ gravity within non-minimally coupled systems is constrained as shown in \cite{22}. Observational constraints on the non-metricity of $f(Q)$ gravity, achieved by modifying the growth of linear perturbations, were discussed in \cite{23, 24}, which accurately recreate the history of the $\Lambda$CDM. Shekh \cite{25} introduced the concept of holographic dark energy (HDE) in the context of $f(Q)$ gravity. The studies referenced in \cite{26,27,28} discuss contemporary cosmic $f(Q)$ gravity models and the underlying factors contributing to the universe's accelerated expansion. Reference \cite{29} demonstrated that the energy conservation criterion aligns with the field equations of the affine connection in $f(Q)$ theory, with the most recent formulations of gravity discussed therein. Jimenez et al. \cite{29a} proposed a new type of gravity, $f(Q)$ gravity, suggesting that the non-metricity factor $Q$ governs gravitational interactions. Lazkoz et al. \cite{29b} used observational constraints to validate their theory's rejection of the $f(Q)$ Lagrangian and its relationship with redshift $z$. Mandal et al. \cite{19} conducted a comprehensive study on energy constraints, which has limited the $f(Q)$ models compatible with the universe's rapid expansion. The $f(Q)$ gravity framework is recognized as the simplest modification of symmetric teleparallel equivalent to general relativity (STEGR), motivating further work within this framework as it addresses various issues \cite{29e,29g,29h}. Mandal et al. \cite{29j,29k} also explored energy constraints in $f(Q)$ theory. Harko et al. \cite{29l} explored the matter coupling phenomenon in modified $f(Q)$ gravity using a power-law function. Additionally, Refs. \cite{29m,29n} examine cosmic acceleration and energy conditions within the framework of teleparallel $f(Q)$ gravity. Several practical applications of modified gravity in different physical scenarios can be found in Refs. \cite{Sharma/2020,Sharma/2018}.

The literature suggests that the conditional probability density $P(\Theta|d)$ can be modeled using a mixture density network (MDNs) when employing a mixture model \cite{29o,29p,29q,29r}. In \cite{29s}, an artificial neural network (ANN) was utilized to estimate parameters and derive the conditional probability density $P(\Theta|d)$, enabling the posterior distribution to be determined at a specific observational data point $d_{0}$. However, learning the conditional probability density $P(\Theta|d)$ becomes challenging when dealing with data that includes covariance. To address this, \cite{29r} proposed using MDNs for parameter estimation within a mixture model framework for $P(\Theta|d)$. The MDN approach also provides a solution to this issue as demonstrated in \cite{29s}. Nonetheless, we discovered that numerous components are necessary, particularly when dealing with values that do not follow a normal distribution. The training time will rise with the use of many components, and there are cases where the MDN becomes unstable during training and fails to establish a posterior distribution.

The structure of this paper is as follows: Section \ref{II} introduces the fundamentals of $f(Q)$ gravity within the framework of the FLRW cosmology. In Section \ref{III}, we outline the methodology for parameter estimation, utilizing 55 observational data points derived from OHD, BAO, and the Pantheon SN Ia compilation, employing both the MCMC approach and deep learning techniques. Section \ref{IV} focuses on the evaluation of key cosmological quantities, including isotropic pressure, energy density, the equation of state (EOS) parameter, and energy conditions. Lastly, Section \ref{V} concludes with a discussion of the results and their implications.

\section{Principles and Implications of $f(Q)$ Gravity}\label{II}
The action for the $f(Q)$ gravity model is expressed as \cite{17}:  
\begin{equation}
S = \int \left[ -\frac{1}{2}f(Q) + L_{m} \right] \sqrt{-g} \, d^{4}x,  \label{eq1}
\end{equation}  
where $f(Q)$, $L_{m}$ and $g$ represent the non-metricity function, the matter Lagrangian, and the determinant of the metric tensor $g_{\mu \nu}$ respectively.  

By varying the action with respect to the metric, the modified field equations are derived as:  
\begin{widetext}
\begin{equation}
\frac{2}{\sqrt{-g}} \nabla_{\gamma} \left( \sqrt{-g} f_{Q} P^{\gamma}{}_{\mu \nu} \right) 
+ \frac{1}{2} f g_{\mu \nu} + f_{Q} \left( P_{\mu \gamma i} Q_{\nu}{}^{\gamma i} 
- 2 Q_{\gamma i \mu} P^{\gamma i}{}_{\nu} \right) = T_{\mu \nu}, \label{eq2}
\end{equation}
\end{widetext}  
Moreover, $f_{Q} = \frac{df}{dQ}$ and $f = cQ^{\gamma}$, where $c$ and $\gamma$ are constants. The variation with respect to the connection leads to the following constraint:  
\begin{equation}
\nabla_{\mu} \nabla_{\gamma} \left( \sqrt{-g} f_{Q} P^{\gamma}{}_{\mu \nu} \right) = 0.  \label{eq3}
\end{equation}  

For this analysis, we assume the FLRW metric, which describes a spatially homogeneous and isotropic universe:  
\begin{equation}
ds^{2} = -dt^{2} + a^{2}(t) \left[ dr^{2} + r^{2} \, d\theta^{2} + r^{2} \sin^{2} \theta \, d\phi^{2} \right], \label{eq4}
\end{equation}  
where $t$ is the cosmic time, $(t, r, \theta, \phi)$ are the co-moving coordinates, and $a(t)$ is the scale factor of the universe.  

In this context, the trace of the non-metricity tensor is given by:  
\begin{equation}
Q = 6H^{2},  \label{eq5}
\end{equation}  
where $H$ is the Hubble parameter.  

If we assume that the fluid is ideal, the stress-energy tensor is as follows:
\begin{equation}
T_{\mu \nu }=\left( \rho +p\right) u_{\mu }u_{\nu }+pg_{\mu \nu },   \label{eq6}
\end{equation}
The four-velocity is denoted as $u_{\mu}$, satisfying the condition $u_{\mu}u^{\mu}=-1$. $\rho$ and $p$ represent the energy density and pressure of the ideal fluid.
By applying the equation of motion (\ref{eq2}) along with the stress-energy tensor (\ref{eq6}) within the FLRW framework, the resulting Friedmann equations for gravity are derived as follows:

\begin{equation}
3H^{2}=\frac{1}{2f_{Q}}\left( \rho +\frac{f}{2}\right) ,     \label{eq7}
\end{equation}
\begin{equation}
\overset{\cdot }{H}+3H^{2}+\frac{\overset{.}{f_{Q}}}{f_{Q}}H=\frac{1}{2f_{Q}}\left( -p+\frac{f}{2}\right) ,  \label{eq8}
\end{equation}
The equation $H=\frac{\overset{\cdot }{a}}{a}$ represents the Hubble parameter, with the overdot denoting the derivative concerning the time variable $t$. \\

The energy density and pressure for the model are given by the following expressions:

\begin{equation}
\rho = \left( 6 H^{2} f_{Q} - \frac{f}{2} \right),  \label{eq9}
\end{equation}

\begin{equation}
p = \frac{f}{2} - 2 f_{Q} \left( \frac{\dot{f_{Q}}}{f_{Q}} H + 3 H^{2} + \dot{H} \right).  \label{eq10}
\end{equation}

\noindent Furthermore, in this paper, the non-metricity function $f(Q) = c\;Q^{\gamma} = c\;6^{\gamma}H^{2\gamma}$, therefore, Eqs. (\ref{eq7}) and (\ref{eq8}) represent a system of two equation with three variables $\rho$, $p$ and $H$. In general, one can not fix its explicit solution without considering i) a well defined relation among the variables $\rho$, $p$ and $H$ or ii) the parameterization of Hubble parameter $H$ in term of red-shift $z$ which is motivated by its ability to capture key cosmological transitions from early deceleration phenomenon to late time accelerating phase and ensures its smooth compatibility with observational data. Therefore, in this study, we reconstruct the $f(Q)$ gravity through a new parameterization of the Hubble parameter, defined as

\begin{equation}
H = \frac{H_0}{\sqrt{2}} \left[ 1 + (1 + z)^{2n} \right]^{\frac{1}{2}},  \label{H}
\end{equation}
where $H_0$ represents the current value of the Hubble constant, and $n$ is an arbitrary constant treated as a free parameter in the derived model of the universe. Furthermore, it is worth to noting that this parametrization of $H$, is all together different from the considered parametrization of $H$ in the literature\cite{Yadav/2024,Koussour/2023FP,Sharma/2025} and it generates $H = H_{0}$ at present epoch ($z = 0$). The ansatz (\ref{H}) reproduced the standard $\Lambda$CDM model ($\Omega_{\Lambda 0} = 0.70$ and $\Omega_{m 0} = 0.30$) as $n \rightarrow \frac{\log \left(\frac{3}{5} (z+1)^3+\frac{2}{5}\right)}{2 \log (z+1)}$ which bounds the values of $n$ from $lim_{z \rightarrow 0}\frac{\log \left(\frac{3}{5} (z+1)^3+\frac{2}{5}\right)}{2 \log (z+1)} = 0.9$ to $lim_{z \rightarrow \infty}\frac{\log \left(\frac{3}{5} (z+1)^3+\frac{2}{5}\right)}{2 \log (z+1)} = 1.5$ that is why we set prior $\{0.9, 1.5\}$ to determine the optimal value of $n$ in the next section.

\noindent In 2004, Sahni et al. \cite{Sahni/2003} have proposed model independent parametrization of $H(z)$ to investigate the statefinder diagnostic of dark energy model. Later on the same $H(z)$ parametrization has been used to discover the dynamics of the universe in $f(Q)$ gravity \cite{Mandal/2022}. Moreover, the various form of Hubble parametrization $H$ in terms of either redshift $z$ or time $t$ are given in Ref. \cite{Pacif/2020}. Motivated by these investigation, in this paper, we configure a new parametrization of $H(z)$ that yields better fitting of model parameter and describes various phases of the universe.\\


\noindent Here, we also describe the relation among the distance modulus $\mu_{th}(z)$, apparent magnitude (m) and absolute magnitude ($M_b$) as following \\
\begin{equation}
\label{D-1}
\mu(z) = -M_{b} + m = \mu_0 + 5log d_L(z),
\end{equation}

\noindent where $d_L(z)$ and $\mu_0$ stand for the luminosity distance and nuisance parameter respectively. Moreover, it is worth noting that the absolute magnitude $M_{b}$ has strong correlation with $H_0$, therefore it is taken as a free parameter.\\

\noindent The $\mu_{0}$ and $d_{L}$ are defined as
\begin{equation}\label{D-2}
\mu_0 = 5log_{10}\left(\frac{H_0^{-1}}{1Mpc}\right) + 25,
\end{equation}
\begin{equation}\label{D-3}
 d_{L} = (1+z)\int^{z}_{0}\frac{H_{0}}{H(x)}dx   
\end{equation} 
Using Eqs. (\ref{D-2}) and (\ref{D-3}) in Eq. (\ref{D-1}), one may obtain
\begin{equation}
\label{D-4}
\mu(z) = 5log_{10}\left(\frac{(1+z)}{1Mpc}\int^{z}_{0}\frac{H_{0}}{H(x)}dx \right) + 25
\end{equation}
 
\begin{equation}\label{mu}
\mu(z) = m_{b} - M = 5log_{10}\left(\frac{(1+z)}{H_{0}}\int^z_0\frac{dx}{h(x)}\right) + \mu_{0} ; \;\; h(x) = \frac{H(x)}{H_{0}}
\end{equation}
where, the zero point offset is $\mu_{0}$. $m_{b}$ and $M$ represent the magnitude and absolute magnitude of distant bright objects respectively. 
\section{Observational constraints}\label{III}
\noindent \textbf{Observational Hubble data (OHD)}: The Cosmic Chronometers (CC) method resulted in $31~H(z)$ measurements derived from CC data and $26~H(z)$ measurements obtained from Baryon Acoustic Oscillation (BAO) data, covering the redshift range $0 \leq z \leq 2.36$. All 57 $H(z)$ data points with its original references \cite{Stern/2010,Simon/2005,Moresco/2012,Zhang/2014,Moresco/2016,h6,h7,h8,h9,h10,h11,h12,h13,h14,h15,h16,h17,h18,h19} are given in Tables I \& II respectively. Furthermore, the CC Dataset's justification based on the differential age development of passively developing galaxies, the CC technique offers a direct, model-independent way to measure the Hubble parameter $H(z)$ at different redshifts \cite{Moresco/2020}. Instead of requiring model-dependent calibrations, CC observations provide a direct probe of the expansion history, in contrast to CMB or BAO approaches, which depend on an assumed background cosmology. We adhere to the technique described in \cite{Moresco/2020} in this study, making sure that both statistical and systematic adjustments are appropriately taken into consideration. Our CC data cover a crucial period in cosmic history, with redshifts ranging from 0.07 to 1.965. In order to address any connections between data points and reduce bias in parameter estimation, we also include the covariance matrix in our chi-squared formalism. We also stress that CC data are derived from extragalactic spectroscopic observations, which achieve great redshift determination accuracy. Accurately estimating stellar population ages, which mostly rely on stellar population synthesis models, is a significant difficulty. These uncertainties have been greatly decreased by recent developments in galaxy evolution models, improving the CC dataset's dependability. This justifies the CC method's sole usage in our research as it continues to be a reliable and independent probe of cosmic expansion.
\begin{table}\label{Tab-2}
\caption{\; The 31 H(z) points obtained from Cosmic Chronometers (CC) method.}
\begin{center}
\begin{tabular}{|c|c|c|c|c|c|c|c|c|c|}
\hline
\textbf{S. N.} & \textbf{z} & \textbf{H(z)}	& \textbf{$\sigma_{i}$} & \textbf{Refs.} & \textbf{S. N.} & \textbf{z} & \textbf{H(z)}	& \textbf{$\sigma_{i}$} &  \textbf{Refs.} \\
\hline
1. & $0.070$ & $69$  & $19.6$ &  \cite{Stern/2010} & 17. & $0.4783$ & $80$ & $99$ &  \cite{Moresco/2016} \\
\hline
2. & $0.090$ & $69$  & $12$ &  \cite{Simon/2005} & 18. & $0.480$ & $97$ & $62$ & \cite{Stern/2010} \\
\hline
3. & $0.120$ & $68.6$  & $26.2$ &  \cite{Stern/2010} & 19. & $0.593$ & $104$ & $13$ &  \cite{Moresco/2012} \\ 
\hline
4. & $0.170$ & $83$  & $8$ & \cite{Simon/2005} & 20. & $0.6797$ & $92$ & $8$ & \cite{Moresco/2012} \\
\hline
5. & $0.1791$ & $75$  & $4$ &  \cite{Moresco/2012} & 21. & $0.7812$ & $105$ & $12$ & \cite{Moresco/2012} \\ 
\hline
6. & $0.1993$ & $75$  & $5$ &  \cite{Moresco/2012} & 22. & $0.8754$ & $125$ & $17$ & \cite{Moresco/2012} \\
\hline
7. & $0.200$ & $72.9$  & $29.6$ &  \cite{Zhang/2014} & 23. &  $0.880$ & $90$ & $40$ & \cite{Stern/2010} \\
\hline
8. & $0.270$ & $77$  & $14$ &  \cite{Simon/2005} & 24. & $0.900$ & $117$ & $23$ & \cite{Simon/2005} \\
\hline
9. & $0.280$ & $88.8$  & $36.6$ &  \cite{Zhang/2014} & 25.  & $1.037$ & $154$ & $20$ & \cite{Moresco/2012} \\  
\hline
10. & $0.3519$ & $83$  & $14$ &  \cite{Moresco/2012} & 26. & $1.300$ & $168$ & $17$ & \cite{Simon/2005} \\
\hline
11. & $0.3802$ & $83$  & $13.5$ &  \cite{Moresco/2016} & 27. & $1.363$ & $160$ & $33.6$ & \cite{h7} \\
\hline
12. & $0.400$ & $95$  & $17$ &  \cite{Simon/2005} & 28.  & $1.430$ & $177$ & $18$ & \cite{Simon/2005} \\ 
\hline
13. & $0.4004$ & $77$ & $10.2$  & \cite{Moresco/2016} & 29. & $1.530$ & $140$ & $14$ & \cite{Simon/2005} \\  
\hline
14. & $0.4247$ & $87.1$ & $11.2$ &  \cite{Moresco/2016} & 30. & $1.750$ & $202$ & $40$ & \cite{Simon/2005} \\
\hline
15. & $0.4497$ & $92.8$ & $12.9$ & \cite{Moresco/2016} & 31. & $1.965$ & $186.5$ & $50.4$ & \cite{h7}\\
\hline
16. & $0.470$ & $89$ & $34$ & \cite{h6} & - & - & - & - & -\\
\hline
 \end{tabular}
\end{center}
\end{table}
\begin{table}\label{Tab-1}
\caption{\; The 26 H(z) points obtained obtained from Baryon Acoustic Oscillation (BAO).}
\begin{center}
\begin{tabular}{|c|c|c|c|c|c|c|c|c|c|}
\hline
\textbf{S. N.} & \textbf{z} & \textbf{H(z)}	& \textbf{$\sigma_{i}$} & \textbf{Refs.} & \textbf{S. N.} & \textbf{z} & \textbf{H(z)}	& \textbf{$\sigma_{i}$} &  \textbf{Refs.} \\
\hline
1. & $0.24$ & $79.69$ & $2.99$ &  \cite{h8} & 14. &  $0.52$ & $94.35$ & $2.64$  & \cite{h10}  \\
\hline
2. & $0.30$& $81.7$ & $6.22$ &  \cite{h9} & 15. & $0.56$ & $93.34$ & $2.3$ &  \cite{h10} \\
\hline
3. &  $0.31$ & $78.18$ & $4.74$ &  \cite{h10} & 16. & $0.57$ & $87.6$ & $7.8$  & \cite{h14}\\
\hline
4. & $0.34$ & $83.8$ & $3.66$ &  \cite{h8} & 17. &  $0.57$ & $96.8$ & $3.4$ & \cite{h15} \\
\hline
5. & $0.35$ & $82.7$ & $9.1$ &  \cite{h11} & 18. & $0.59$ & $98.48$ & $3.18$ &  \cite{h10} \\ 
\hline
6. & $0.36$ & $79.94$ & $3.38$ &  \cite{h10} & 19. & $0.60$ & $87.9$ & $6.1$ &  \cite{h13} \\ 
\hline
7. & $0.38$ & $81.5$ & $1.9$ &  \cite{h12} & 20. & $0.61$ & $97.3$ & $2.1$ &  \cite{h12} \\ 
\hline
8. & $ 0.40$ & $82.04$ & $2.03$ &  \cite{h10} & 21. & $0.64$ & $98.82$ & $2.98$ &  \cite{h10}\\
\hline
9. & $0.43$ & $86.45$ & $3.97$ &  \cite{h8} & 22. & $0.73$ & $97.3$ & $7.0$ & \cite{h13}\\
\hline
10. & $0.44$ & $82.6$ & $7.8$ &  \cite{h13} & 23. & $2.30$ & $224$ & $8.6$ &  \cite{h16}\\
\hline
11. & $0.44$ & $84.81$ & $1.83$ & \cite{h10} & 24. & $2.33$ & $224$ & $8$ & \cite{h17} \\
\hline
12. & $0.48$ & $87.79$ & $2.03$ & \cite{h10} & 25. & $2.34$ & $222$ & $8.5$ & \cite{h18}  \\
\hline
13. & $0.51$ & $90.4$ & $1.9$ &  \cite{h12} & 26. & $2.36$ & $226$ & $9.3$ & \cite{h19}\\
\hline 
\end{tabular}
\end{center}
\end{table}

\noindent\textbf{Type Ia supernovae (SNIa) Pantheon sample}: A recent addition to this collection is the updated Pantheon dataset using the Pantheon sample \cite{Scolnic/2018}. 

\subsection{Bayesian Analysis}
\noindent In the following subsection, we impose constraints on these parameters using the latest observational datasets to ensure the model's consistency with observation data. To achieve this, various cosmological observations can be utilized, including Hubble parameter \( H(z) \) measurements, Pantheon plus compilation of type Ia Supernovae (SNe Ia) data. In this study, we adopt a Bayesian statistical approach and employ the Markov Chain Monte Carlo (MCMC) method to derive the posterior distributions of the model parameters \cite{57}. The MCMC analysis is performed using the widely used emcee package \cite{58}, ensuring robust parameter estimation. We assume uniform prior distribution for all free parameters, and their ranges in the MCMC fitting process are set to be as following: $H_{0} \in \{60, 80\}$ \& $n \in \{0.9, 1.5\}$. 

\noindent Among the available datasets, we specifically incorporate the Hubble parameter measurements, often referred to as the \( H(z) \) dataset. These measurements play a crucial role in constraining cosmological models, as they provide insights into the expansion history of the universe. The best-fit values of the parameters are determined by maximizing the likelihood function, which is expressed as 
\begin{equation}\label{19}
\mathcal{L} \propto \exp\left(-\chi^2/2\right)
\end{equation}
\noindent where $ \chi^2 $ represents the chi-square function used to assess the goodness of fit between the theoretical predictions and observational data. In this section, we focus on estimating the free parameters of the model using observational data, achieved through the application of $\chi^2$-minimization techniques. The model parameters $n$ and $H_0$ are determined using the following methodologies:
\begin{figure}[H]
\centering
\includegraphics[width = 0.48\textwidth]{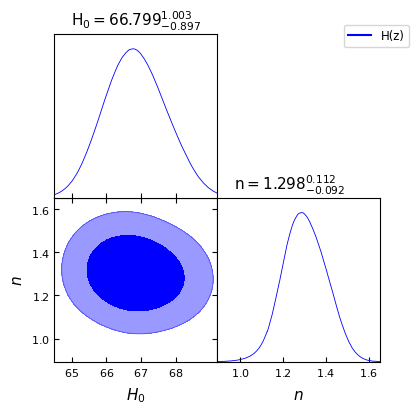}
\includegraphics[width = 0.48\textwidth]{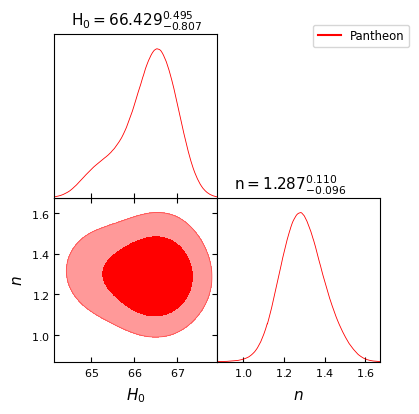}
\caption{The two-dimensional contours depicting the $1\sigma$ and $2\sigma$ confidence regions.}\label{F1}
\end{figure}
\begin{figure}[H]
\centerline{\includegraphics[width = 0.60\textwidth]{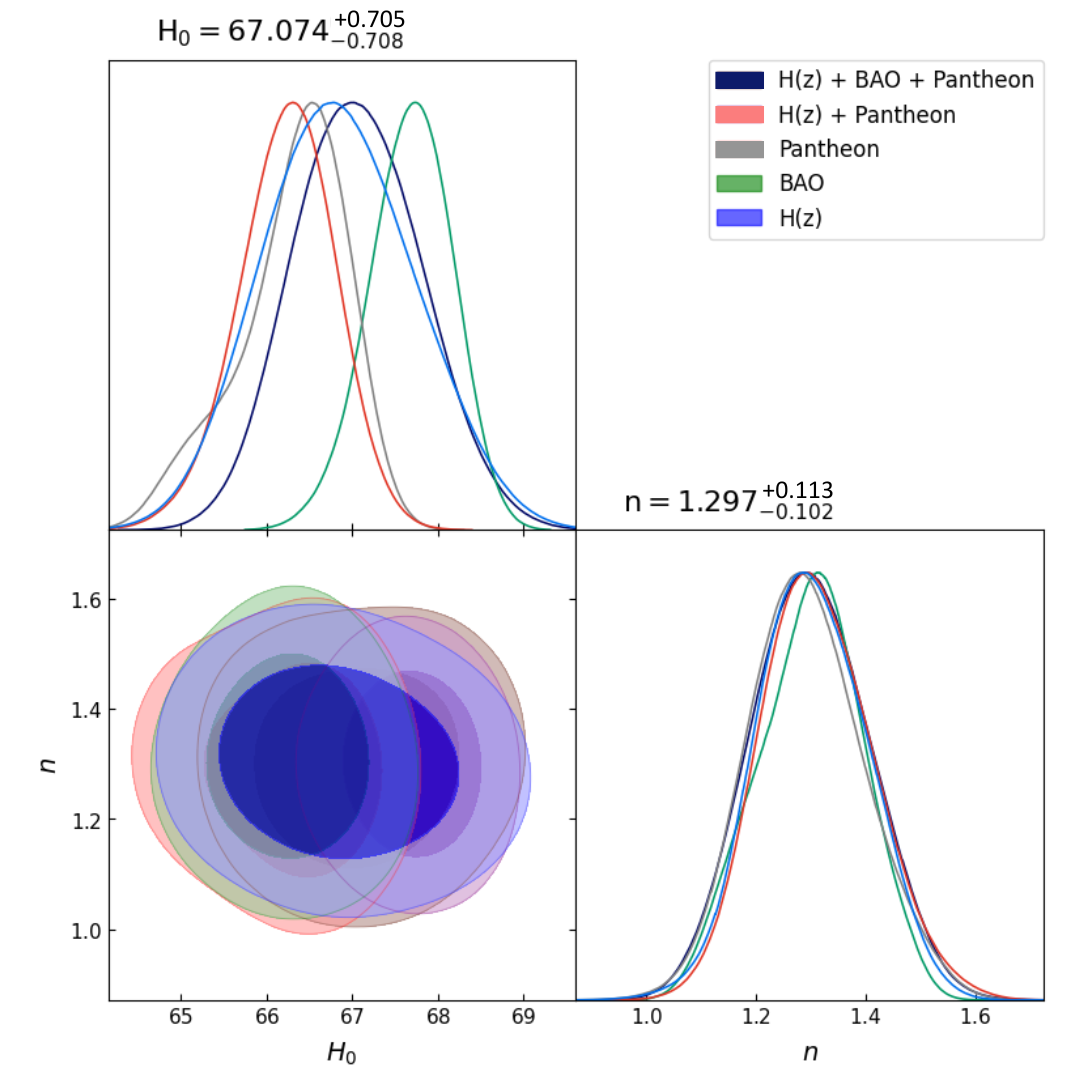}}
\caption{The integrated representation of two-dimensional contours at the $1\sigma$ and $2\sigma$ confidence levels.}\label{F2}
\end{figure}

\noindent For the purpose of evaluating the model parameters $n$ and $H_{0}$ of this model, observational data and statistical approaches may be used. Let $E_{obs}$ represent the values that have been observed and let $E_{th}$ represent the values that have been theoretically calculated to conform to the model of the universe. Estimating the parameters of the model can be achieved by using statistical techniques to compare the two values $E_{obs}$ and $E_{th}$ within the model. For the purpose of obtaining the most accurate estimated values, we used the $\chi^2$ estimator. The symbol $\sigma$ represents the standard error in the values observed. It is possible to interpret the formulation of the $\chi^2$ estimator as
\begin{equation}
\label{chi1}
\chi^{2} = \sum_{i=1}^{N}\left[\frac{E_{th}(\Phi_{c},z_{i})- E_{obs}(\Phi_{l}z_{i})}{\sigma_{i}}\right]^{2}
\end{equation}
In this context, the symbols $E_{th}(\Phi_{c},z_{i})$ and $E_{obs}(\Phi_{l},z_{i})$ represent the theoretical values and the observed values of the relevant parameters by themselves. The $\Phi_{c}$ is the cosmological parameter sets, and $\Phi_{c} = (H_{0}, \mu, n)$. The $\Phi_{l}$ denotes the nuisance parameter sets and we have $\phi_{l} = (H_{0}, n)$ and $\Phi_{l} = (\mu, n)$ for OHD and SN Ia Pantheon sample respectively. Further, we refer to the standard errors in $E_{obs}(z_{i})$ as $\sigma_{i}$ and the number of data points as N.\\

\noindent Figure \ref{F1} shows the two-dimensional confidence contours at $1\sigma$ and $2\sigma$ intervals, which constrain our model using the OHD dataset, comprising both the BAO and Pantheon SN Ia data. In contrast, Figure \ref{F2} presents a combined view of the $1\sigma$ and $2\sigma$ confidence regions. The Hubble constant, $H_0$, is expressed in $\text{km}\;\text{s}^{-1}\;\text{Mpc}^{-1}$. The values estimated for $H_0$ and $n$ are as follows: $H_0 = 67.799^{+1.003}_{-0.897}\;\text{km}\;\text{s}^{-1}\;\text{Mpc}^{-1}$, $n = 1.298^{+0.112}_{-0.092}$ for the OHD dataset, and $H_0 = 66.429^{+0.495}_{-0.807}\;\text{km}\;\text{s}^{-1}\;\text{Mpc}^{-1}$, $n = 1.287^{+0.110}_{-0.096}$ when both the OHD and Pantheon SN Ia data are used for the constraints.
\subsection{Neural Networks}

\noindent In this study, the term \textit{deep learning} specifically refers to the use of Artificial Neural Networks (ANNs), Mixture Density Networks (MDNs), and a hybrid Mixed Neural Network (MNN) architecture.

\noindent In this subsection, we adopt the \texttt{CoLFi} package \cite{29r}  for simulation-based cosmological parameter inference. Furthermore, it is worthwhile to note that here, we employ \texttt{CoLFi} as a tool rather than develop it, therefore, we refer a detailed comparison of its advantages over other methods to the original works by Wang et al \cite{29r,29s}. The loss function of neural networks methods are read as follows:

\subsubsection{Artificial Neural Networks}


\noindent In Ref. \cite{NIPS2017_addfa9b7}, it is recommended for automatic neural networks (ANN) to employ absolute deviations as their loss function. Therefore, the loss function for ANN method is read as


\begin{equation}
\mathcal{L} = \mathbb{E} \left( \frac{1}{N} \sum_{i=1}^{N} \left| \theta_i - \hat{\theta}_i \right| \right),
\end{equation}
The parameter space point \( \theta \) represents the estimated parameters, while the symbol \( N \) denotes the number of cosmological parameters and the target is symbolized by \( \hat{\theta} \) in the training set. 


\begin{figure}[htbp]
\centering
\includegraphics[width=0.7\textwidth]{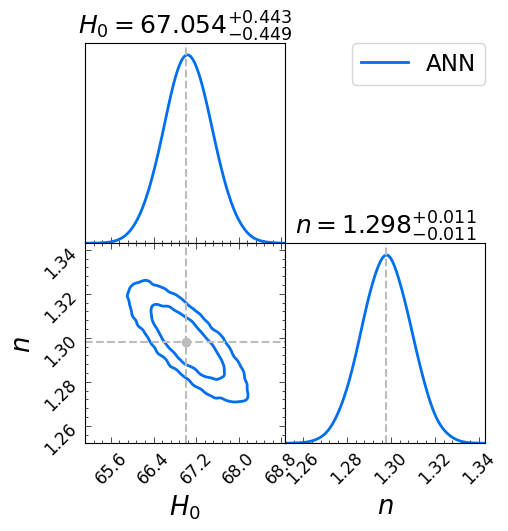}
\caption{The 2D contour and marginalized distributions of $H_0$ and $n$, displaying the contours of 1$\sigma$ and 2$\sigma$.}
\label{fig:sub2}
\end{figure}

\subsubsection{Mixture Density Network}

\noindent The loss function of MDNs methods is read as

\begin{equation}
\mathcal{L} = \mathbb{E} \left[ -\ln \left( \sum_{i=1}^{K} \omega_i \cdot \frac{\exp \left( -\frac{1}{2} (\hat{\theta} - \mu_i)^\top \Sigma_i^{-1} (\hat{\theta} - \mu_i) \right)}{\sqrt{(2\pi)^N |\Sigma_i|}} \right) \right], \label{eqn:loss_multi}
\end{equation}
where $\hat{\theta}$ (or $\hat{\boldsymbol{\theta}}$) represents the estimated parameters in both the single- and multi-parameter cases. 



\subsubsection{Mixture Neural Network}



\noindent The modified loss function for model with multiple parameters is obtained as

\begin{equation}
\mathcal{L} = \mathbb{E} \left[ - \ln \left( \sum_{i=1}^{K} \omega_i \cdot \frac{\exp \left( -\frac{1}{2} (\boldsymbol{\theta} - \hat{\boldsymbol{\theta}})^\top \boldsymbol{\Sigma}_i^{-1} (\boldsymbol{\theta} - \hat{\boldsymbol{\theta}}) \right)}{(2\pi)^{n/2} |\boldsymbol{\Sigma}_i|^{1/2}} \right) \right], \label{eqn:loss_multi}
\end{equation}

\noindent In this setup, the cosmological parameters $\theta$ (or $\boldsymbol{\theta}$) are considered as specific points within the parameter space. The target parameter values, $\hat{\theta}$ (or $\hat{\boldsymbol{\theta}}$), are derived from the training set, while $\boldsymbol{\omega}$ represents the mixture weights. 


\noindent Using the modified loss function, we can compute the posterior distribution directly, which is analogous to the output of the ANNs. Additionally, the performance and results of the deep learning-based neural network, utilizing the COLFi methodology \cite{29r,29s}, applied to the 57-point $H(z)$ data set are visualized in Figs. 3 - 11.



\noindent In Fig. \ref{FR1}, we present a comparison of our model with the $\Lambda$CDM paradigm. This is achieved through the presentation of $H(z)$ and $\mu(z)$ plots, which utilize the MCMC and Deep Learning methodologies. This plots demonstrates the model's ability to replicate observational trends, offering a quantitative evaluation of its effectiveness. Although a formal Bayesian evidence comparison is not within the scope of this analysis, the results indicate that the model provides a statistically comparable fit to the observational data. Moreover, in modified gravity circumstances, likelihood functions may display non-Gaussian distributions with degenerate zones, despite the fact that our model only has two free parameters. Complex posteriors may be difficult for traditional Bayesian techniques to handle, thus simulation-based inference is a potent substitute. As shown in \cite{29r}, our use of neural density estimation allows for a more flexible exploration of parameter space. Furthermore, it is worthwhile to note that the key advantage of the \texttt{COLFi} approach, compared to the MCMC method, is its ability to avoid the need for likelihoods, demonstrating its efficiency in cosmological parameter estimation.

\noindent While the overall agreement between the Bayesian (MCMC) and deep learning (ANN, MDN, MNN) methods is strong, particularly at low to intermediate redshifts, a slight divergence is observed at higher redshift values. This difference, although within the 1$\sigma$ error bounds, arises due to the subtle variation in the best-fit values of $n$ and $H_0$ obtained from the two approaches. Since the redshift evolution of $H(z)$ is highly sensitive to these parameters at larger $z$, even small discrepancies can become amplified in that regime. Additionally, the CoLFI-based neural network methods are trained on synthetic data generated across a sampled hyperellipsoid, which can introduce mild smoothing or regularization effects not present in the purely data-driven MCMC approach. These factors collectively contribute to the observed high-redshift divergence, and do not indicate a breakdown of consistency but rather reflect methodological differences in how the posterior distributions are learned. 

\noindent While both the MCMC-based Bayesian approach and the simulation-based neural network framework yield consistent constraints on the cosmological parameters, Neural Network Method offers several key advantages. Firstly, it circumvents the explicit computation of likelihood functions, which can be computationally expensive or even intractable in modified gravity models with complex dynamics. Secondly, the CoLFI framework enables efficient exploration of high-dimensional and potentially degenerate posterior distributions using neural density estimators, which adapt more flexibly to non-Gaussian features. Unlike traditional MCMC methods that may require long chains and careful convergence diagnostics, neural networks - once trained - allow for instantaneous posterior evaluation, making them particularly suitable for repeated or real-time inference tasks. Moreover, the hybrid architecture of the Mixture Neural Network (MNN) integrates the strengths of both ANN and MDN, leading to stable training and more accurate uncertainty quantification. These attributes collectively make Method B a robust and scalable alternative, especially for cosmological models beyond the standard $\Lambda$CDM framework.

\noindent Furthermore, this quantitative analysis shows that Neural Network Method provides the best balance between accuracy and efficiency. Better comparison about MCMC versus Deep Learning with CPU usage and runtime is provided in Wang et al. \cite{29r}.

\section{Physical parameters of the model}\label{IV}
\subsection{\textbf{The pressure and energy density}}
\noindent The pressure and energy density in this universe model are computed as
\begin{equation}
\rho = 2^{(-1 + \gamma)} 3^{\gamma} \left(1 + (1 + z)^{2n} H_{0}^{2}\right)^{\gamma} \left(-1 + 12 \left(1 + (1 + z)^{2n} H_{0}^{2}\right)\right)
   \label{eq10}
\end{equation}
\begin{equation}
p=\frac{1}{2} \left( 48 n \left((1 + z)^{2n} + (1 + z)^{4n}\right) H_{0}^{4} + 6^{\gamma} \left(1 + (1 + z)^{2n} H_{0}^{2}\right)^{\gamma} \left(1 + 4 \left(-3 + (-3 + n)(1 + z)^{2n}\right) H_{0}^{2}\right) \right)
 \label{eq10p}
\end{equation}

\noindent The general expression for $f(Q) = cQ^{\gamma}$, from which Equations (\ref{eq10}) and (\ref{eq10p}) are derived, reveals that $c = 1$ and parameter $\gamma$ determine its various form. By analyzing specific values of $\gamma$, one can obtain a distinct form of $f(Q)$, which plays a crucial role in the model's behavior. When $\gamma = 1$, the function $f(Q)$ takes a linear form; for $\gamma = 2$, it becomes quadratic; and for $\gamma = 3$, the function is cubic. The evolution of the energy density $\rho$ and pressure $p$ as functions of the redshift $z$ for the cases $\gamma = 1$, $\gamma = 2$, and $\gamma = 3$ are illustrated in Figures \ref{gamma1}, \ref{gamma2}, and \ref{gamma3}, respectively. \textcolor{blue}{It is important to note that both the energy density \( \rho \) and pressure \( p \) are treated as dimensionless quantities in this analysis, as they are expressed in normalized form with respect to the critical density and natural units where \( 8\pi G = c = 1 \).}
\begin{figure}[H]
\centering
\includegraphics[width = 0.48\textwidth]{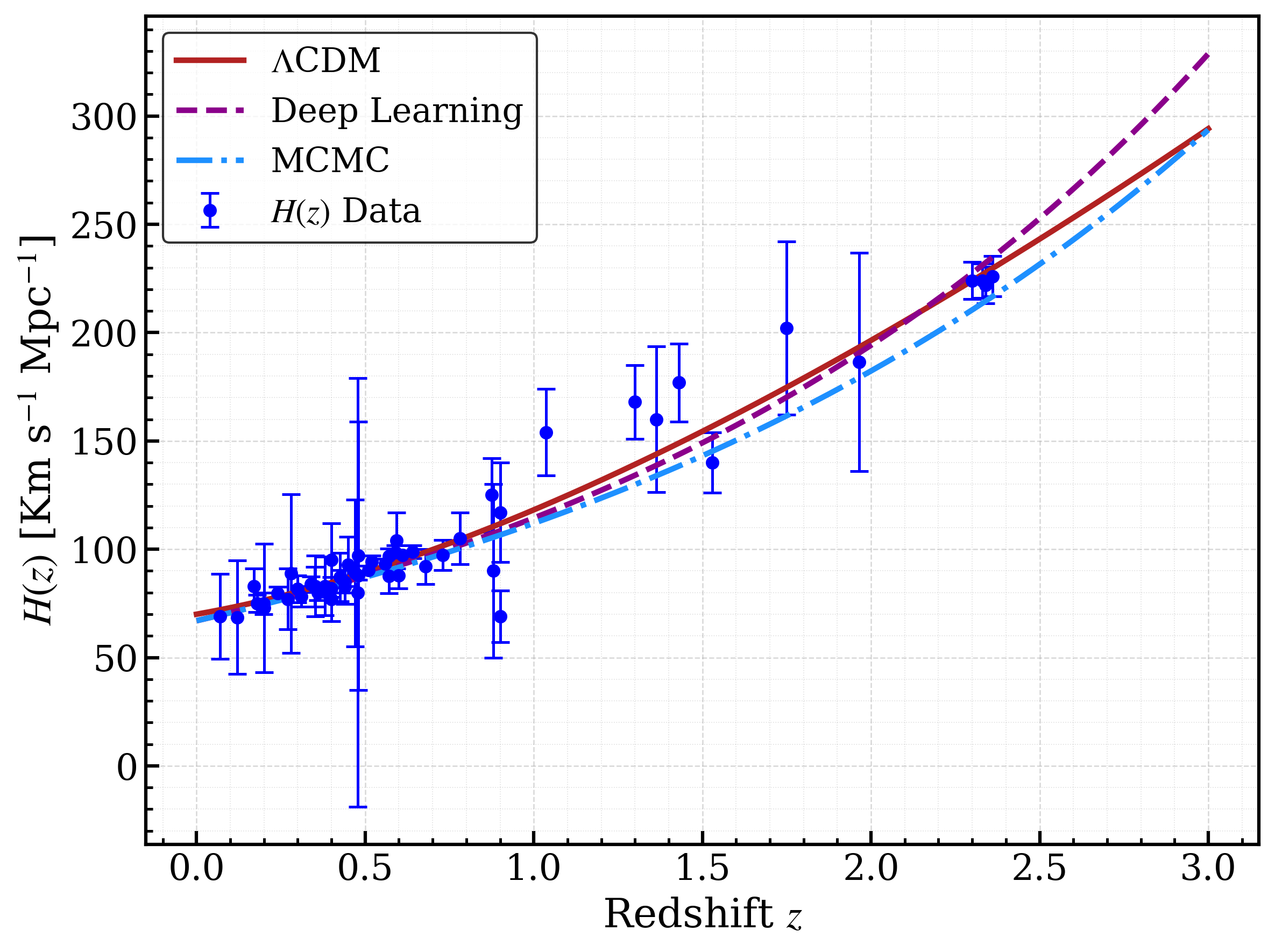}
\includegraphics[width = 0.48\textwidth]{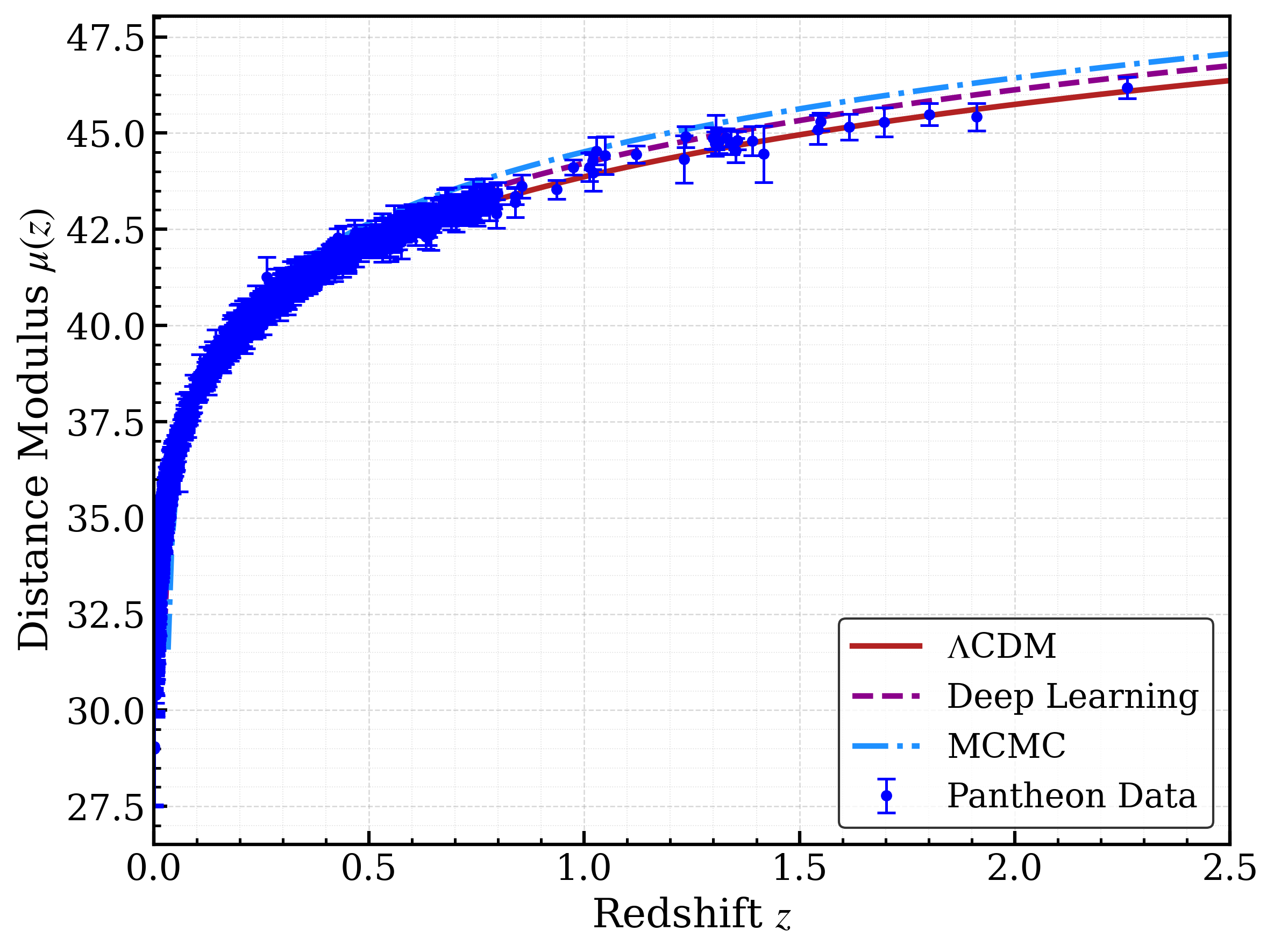}
\caption{The left panel of above figure shows the variation of H(z) of our model with redshift z and its comparison with $\Lambda$CDM model for OHD data while the right panel of above figure exhibits the variation of distance modulus $\mu(z)$ of our model with redshift z and its comparison with $\Lambda$CDM model for Pantheon sample of SN Ia data.}\label{FR1}
\end{figure}
\begin{figure}
\centering
\includegraphics[scale=0.35]{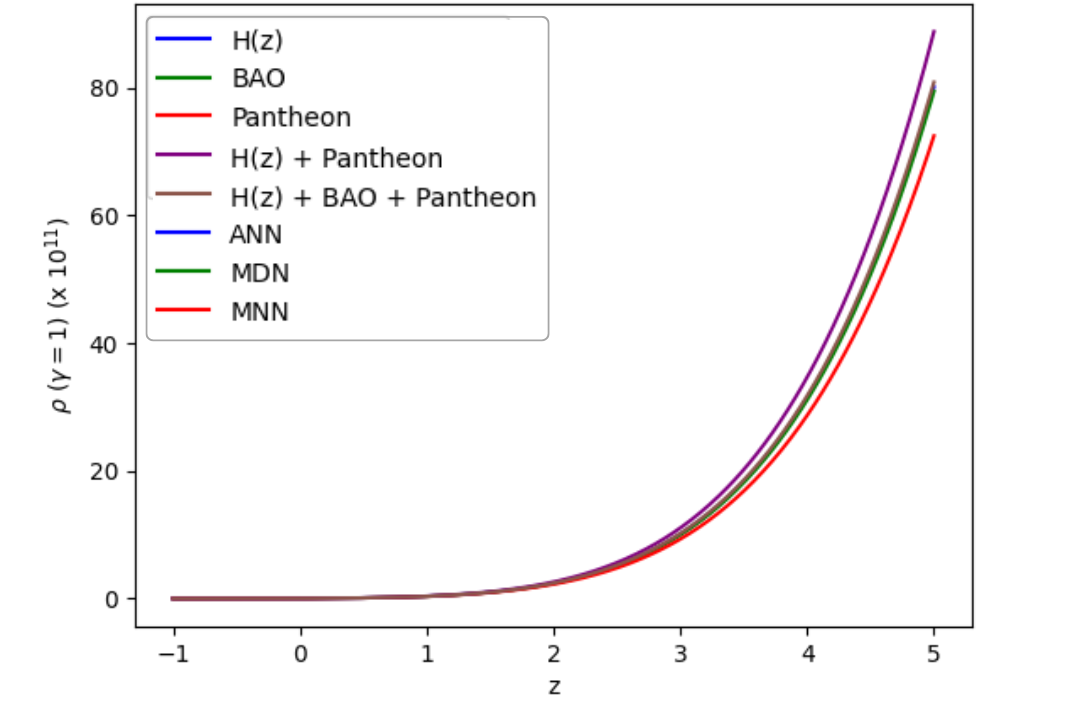}
\includegraphics[scale=0.35]{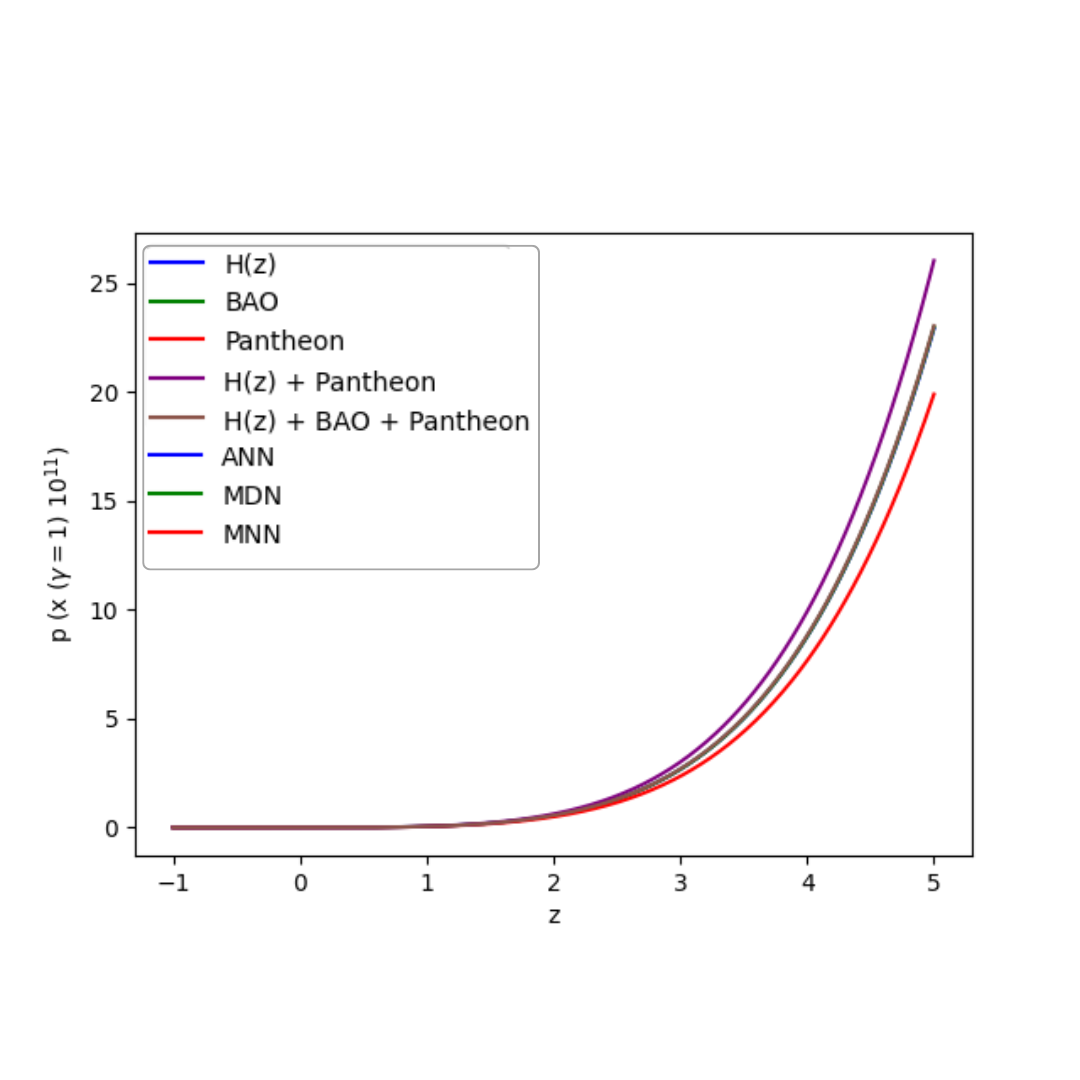}
\caption{For Bayesian and Deep Learning Statistics, the development of energy density $\rho$ and pressure $p$ with redshift $z$ for $\gamma = 1$.}\label{gamma1}
\end{figure}
\begin{figure}
\centering
\includegraphics[scale=0.35]{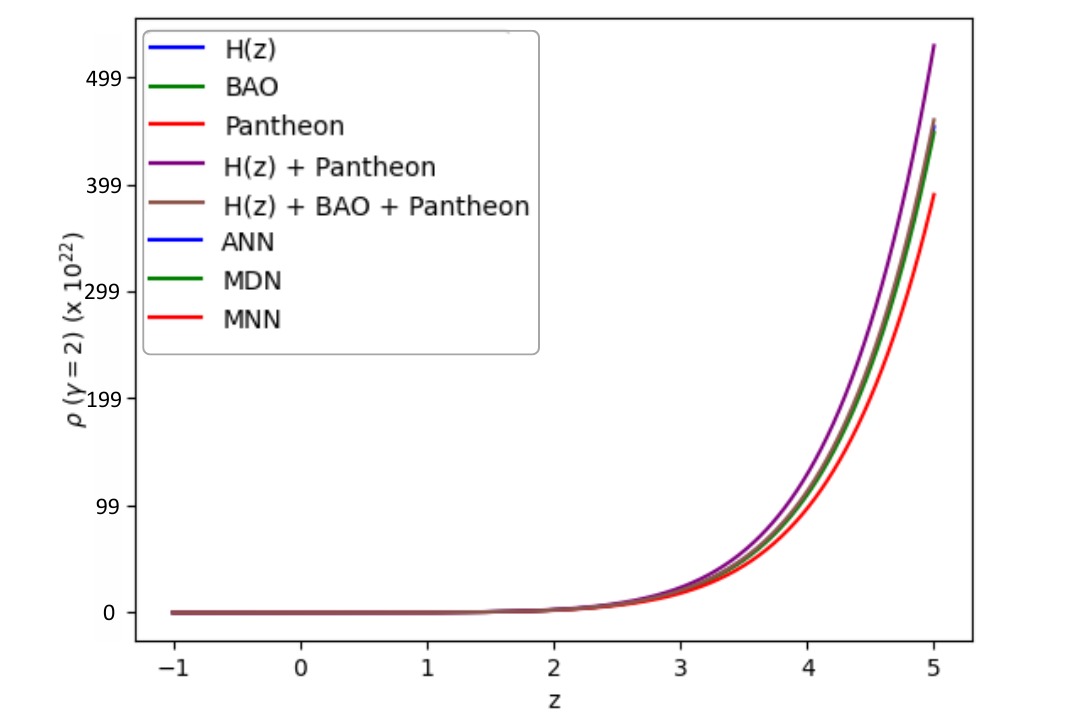}
\includegraphics[scale=0.35]{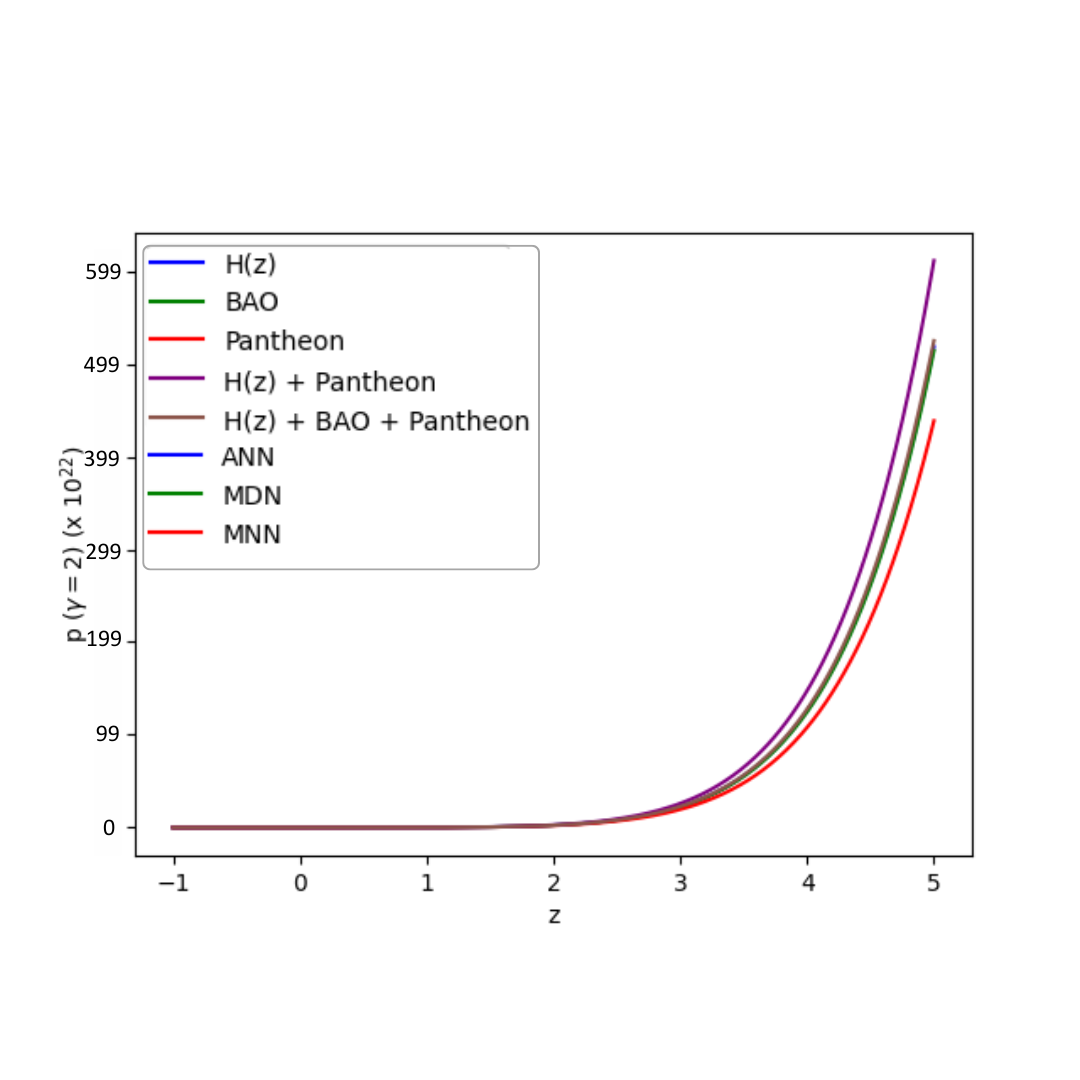}
\caption{For Bayesian and deep learning statistics, the development of energy density $\rho$ and pressure $p$ with redshift $z$ for $\gamma = 2$.}\label{gamma2}
\end{figure}
\begin{figure}
\centering
\includegraphics[scale=0.35]{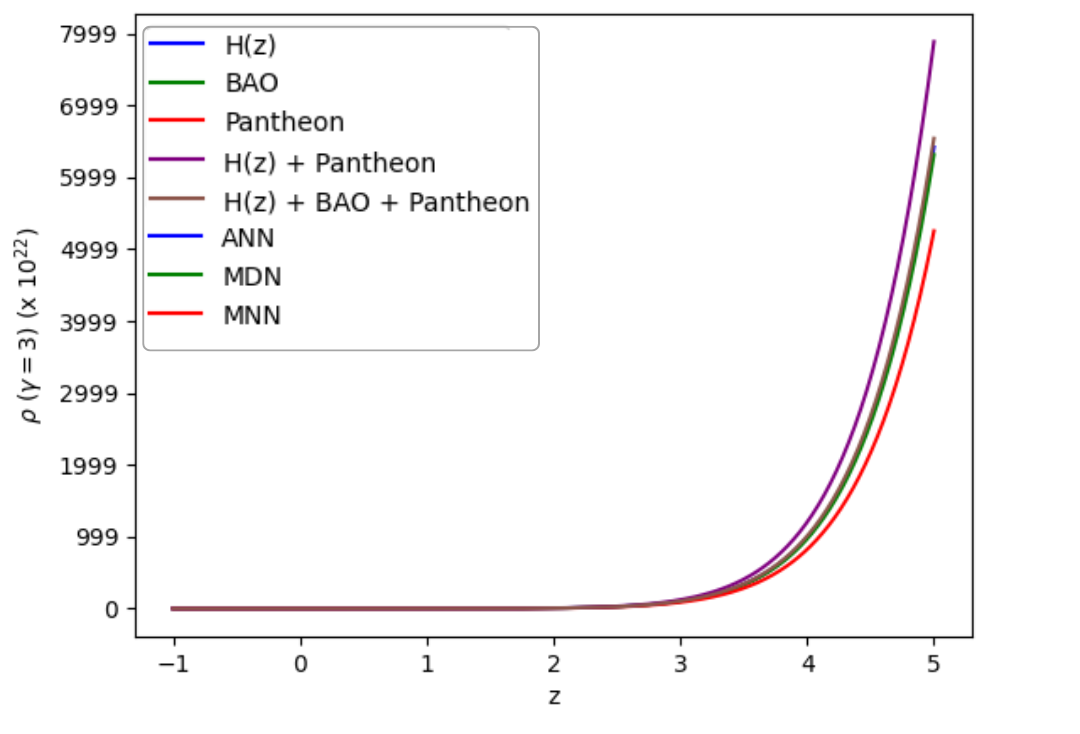}
\includegraphics[scale=0.35]{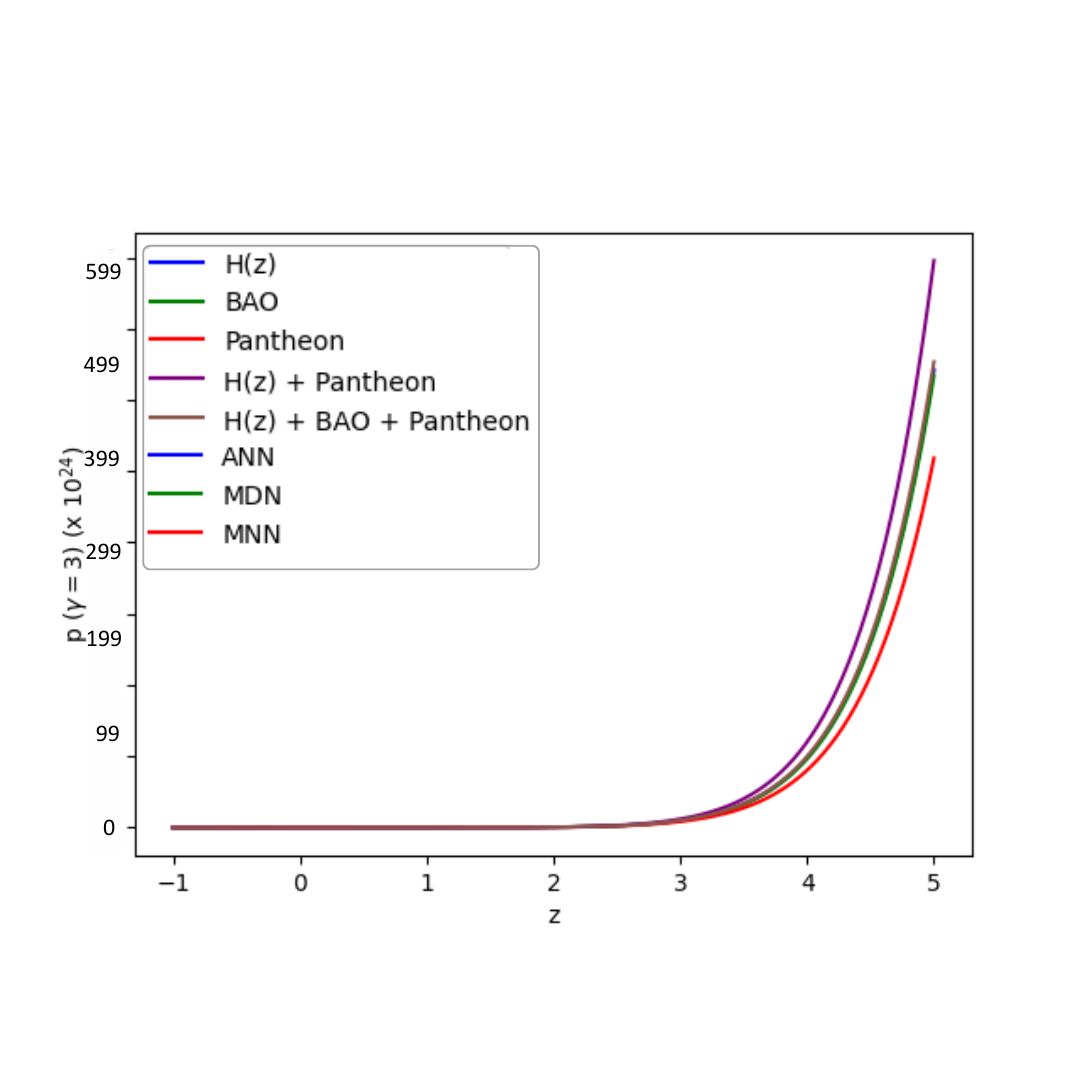}
\caption{For Bayesian and deep learning statistics, the development of energy density $\rho$ and pressure $p$ with redshift $z$ for $\gamma = 3$.}\label{gamma3}
\end{figure}

\noindent Moreover, Solanki et al. \cite{Solanki/2021} explored a linear model within the context of $f(Q)$ gravity to shed light on the late-time acceleration of the universe, which is attributed to the dynamics of viscous fluids. Building upon this, Koussour et al. \cite{Koussour/2022} examined the model further by considering the impact of varying expansion rates. In their subsequent work \cite{Koussour/2022a}, the authors extended their analysis to the quadratic form of $f(Q)$ gravity, investigating transitional phenomena that emerge from a hybrid expansion framework. A novel assessment of finite-time cosmological singularities and the possible future evolution of the universe is presented in Ref. \cite{Haro/2023}. Additional contributions to the field of $f(Q)$ gravity are found in the works of Nojiri et al. \cite{Nojiri/2024}, Hu et al. \cite{Hu/2023}, and Capozziello et al. \cite{Capozziello/2024}. In this study, we reconstruct the $f(Q)$ theory of gravity assuming $f(Q) = cQ$. The functional form of $f(Q)$ encompasses linear, quadratic, cubic, and bi-quadratic cases, as indicated by parameter $\gamma$. In the works of Solanki et al. \cite{Solanki/2021}, Koussour et al. \cite{Koussour/2022}, and Koussour et al. \cite{Koussour/2022a}, the authors specifically investigated the linear and quadratic forms of $f(Q)$ gravity. Thus, the $f(Q)$ gravity model derived in this study is more general and not confined to previously specified models.

\noindent The upper panels of Figs. \ref{gamma1}, \ref{gamma2} \& \ref{gamma3} illustrate the evolution of the energy density $\rho$ as a function of redshift $z$ in the model developed in this study. At $z = 0$, corresponding to the present epoch of the universe, the energy density attains a small positive value. As $z$ increases, the energy density decreases, reaching a moderate positive value. On the other hand, the lower panel of Figs. \ref{gamma1}, \ref{gamma2} \& \ref{gamma3} depict the relationship between the pressure $p$ and redshift $z$. It shows that the pressure decreases as $z$ decreases, reflecting the changing dynamics of the universe over its evolving process.
\newpage
\subsection{The equation of state (EOS) parameter}
\noindent The equation governing the EOS parameter $\omega$ for the cubic $f(Q)$ gravity model is expressed as follows:

\begin{equation}
\omega = \frac{6^{-\gamma} \left(1 + (1 + z)^{2n} H_{0}^{2}\right)^{-\gamma} \left(48 n \left((1 + z)^{2n} + (1 + z)^{4n}\right) H_{0}^{4} + 6^{\gamma} \left(1 + (1 + z)^{2n} H_{0}^{2}\right)^{\gamma} \left(1 + 4 \left(-3 + (-3 + n)(1 + z)^{2n}\right) H_{0}^{2}\right) \right)
}{
-1 + 12 \left(1 + (1 + z)^{2n}\right) H_{0}^{2}
}   \label{eq10omega}
\end{equation}
\begin{figure}
\centering
\includegraphics[scale=0.35]{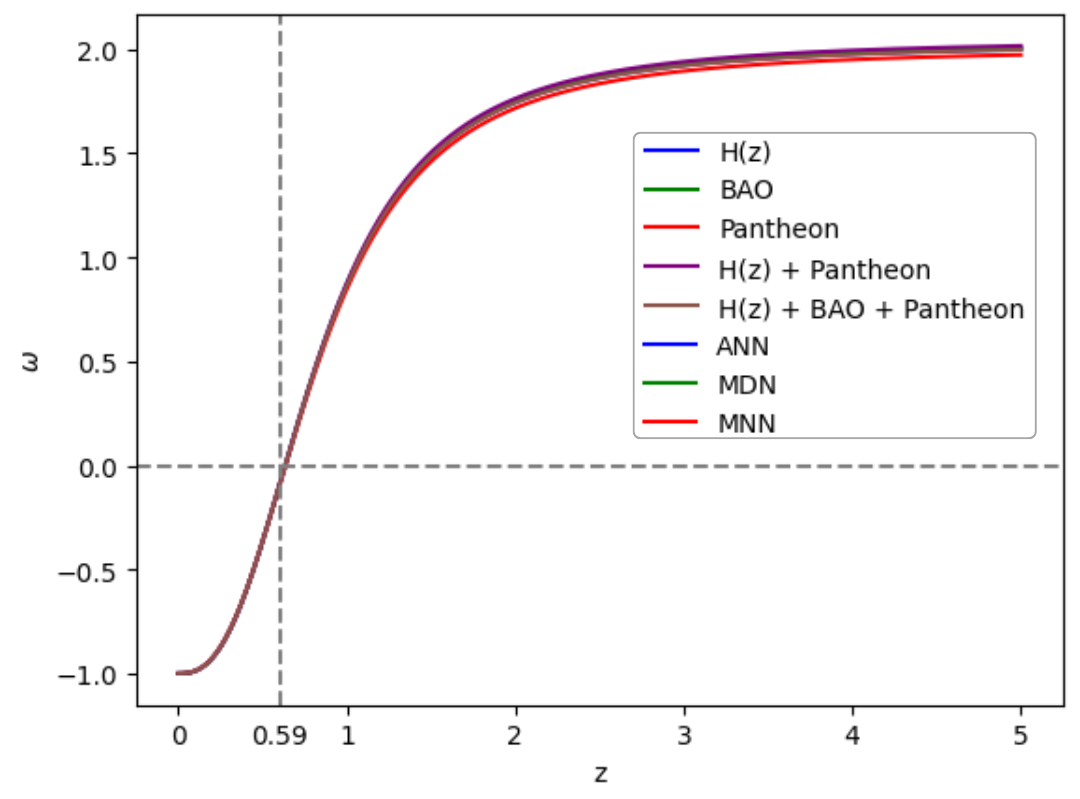}
\caption{The evolutionary behavior of the EOS parameter $\omega$ with regard to the redshift $z$ using Bayesian (left) and deep learning statistics (right).}\label{eos}
\end{figure}
\noindent Figure \ref{eos} illustrates the evolution of the equation of state (EOS) parameter $\omega$ as a function of redshift $z$ based on our proposed model. It is observed that $\omega$ remains positive for $z \geq 0.59$ and transitions to negative values for $z < 0.59$. The negative $\omega$ values signify an accelerating phase of the universe. Furthermore, the EOS parameter approaches $\omega = -1$ as $z$ approaches $-1$. The behavior of our model aligns with the $\Lambda$CDM framework \cite{Komatsu/2009, Kumar/2011a, Kumar/2011b} at later stages.
\subsection{\textbf{Deceleration parameter}}
\begin{figure}[H]
\centering
\includegraphics[scale=0.7]{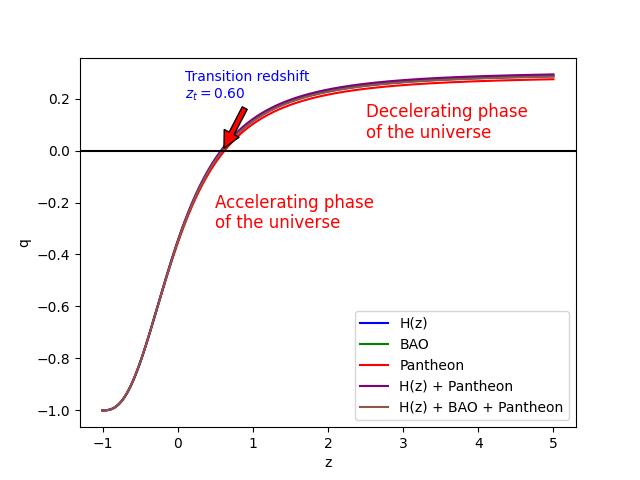}
\caption{The diagram depicts the correlation between $q(z)$ and redshift $z$ through different dataset combinations.}\label{q}
\end{figure}
\noindent The expression for deceleration parameter $q(z)$ is read as
\begin{equation} \label{eq24}
q(z)=-1+\frac{d}{dt}\left(\frac{1}{H}\right) .
\end{equation}
Thus, using Eq. (\ref{H}) in Eq. (\ref{eq24}), we compute $q(z)$ as follows 
\begin{equation} \label{eq25}
q(z) = -\frac{1 + (1+z)^{2n}(1 - n)}{1 + (1+z)^{2n}}
\end{equation}

\noindent Figure \ref{q} depicts the variation in the acceleration-deceleration phases along the redshift axis. Using the OHD and Pantheon SN Ia datasets, the transition redshift value is determined as $z_{t} = 0.60$, consistent with findings reported in Refs. \cite{30,31}. Recent studies have updated this transition redshift to $z_{t} = 0.69^{+0.23}_{-0.12}$. Additionally, Lu \cite{Lu/2011} calculated $z_{t} = 0.60^{+0.21}_{-0.12}$, while Yang \cite{Yang/2020} provided a value of $z_{t} = 0.723^{+0.34}_{-0.16}$, referenced further in Yadav et al.'s work \cite{Yadav/2021prd}. The transition redshift obtained in this study closely aligns with the value presented in Ref. \cite{Yang/2020}.

\noindent According to the proposed model, Figure \ref{q} demonstrates that the universe undergoes an accelerating phase for redshift values $z < z_{t}$ and a decelerating phase for $z > z_{t}$. Notably, at $z = -1$, the parameters $q = -1$ and $\frac{dH}{dt} = 0$ indicate the universe's maximum expansion rate. Our analysis also reveals that the current universe is undergoing rapid expansion, with the deceleration parameter taking the value $q_{0} = -0.3527$. For additional constraints on $q(z)$, relevant discussions can be found in Refs. \cite{Yadav/2021prd,Capozziello/2020}.

\subsection{\textbf{Statefinder analysis}}

\noindent The statefinder parameters $(r, s)$ are commonly used to distinguish between various Dark Energy (DE) models. These parameters are defined as:

\begin{equation}    \label{eq26}
r = q(z) + 2q^{2}(z) - H^{-1}\dot{q},
\end{equation} 
\begin{equation}\label{eq27}
s = \frac{(r - 1)}{3(q - 1/2)}.
\end{equation} 

\noindent Different DE models can be identified using specific values of $(r, s)$. For example: $(r = 1, s = 0)$ corresponds to the $\Lambda$CDM model, $(r > 1, s < 0)$ is indicative of the CG model, and $(r < 1, s > 0)$ characterizes the quintessence model. These models are explored in sequence.

\noindent By integrating Equations (\ref{H}), (\ref{eq26}), (\ref{eq27}), and (\ref{eq24}), the statefinder parameters $(r, s)$ are calculated as follows:

\begin{equation}    \label{eq28}
r = \frac{(-1 + (-1 + n) (1 + z)^{2n}) (-1 - (1 + z)^{2n} + \sqrt{1 + (1 + z)^{2n}} (-1 + (-1 + 2n) (1 + z)^{2n}) H_0)}{((1 + (1 + z)^{2n})^{5/2} H_0)},
\end{equation}
\begin{equation}    \label{eq29}
s = \frac{1}{3} \left( n \left(2 - \frac{2}{1 + (1 + z)^{2n}}\right) + \frac{2 - 2 (-1 + n) (1 + z)^{2n}}{\sqrt{1 + (1 + z)^{2n}} (-3 + (-3 + 2n) (1 + z)^{2n}) H_0} \right).
\end{equation}

\noindent The derivative $\dot{q}$ is expressed as $\dot{q} = \frac{d}{dt}q = -(z+1)H(z)\frac{dq}{dz}$.

\begin{figure}[H]
.\centerline{\includegraphics[scale=0.35]{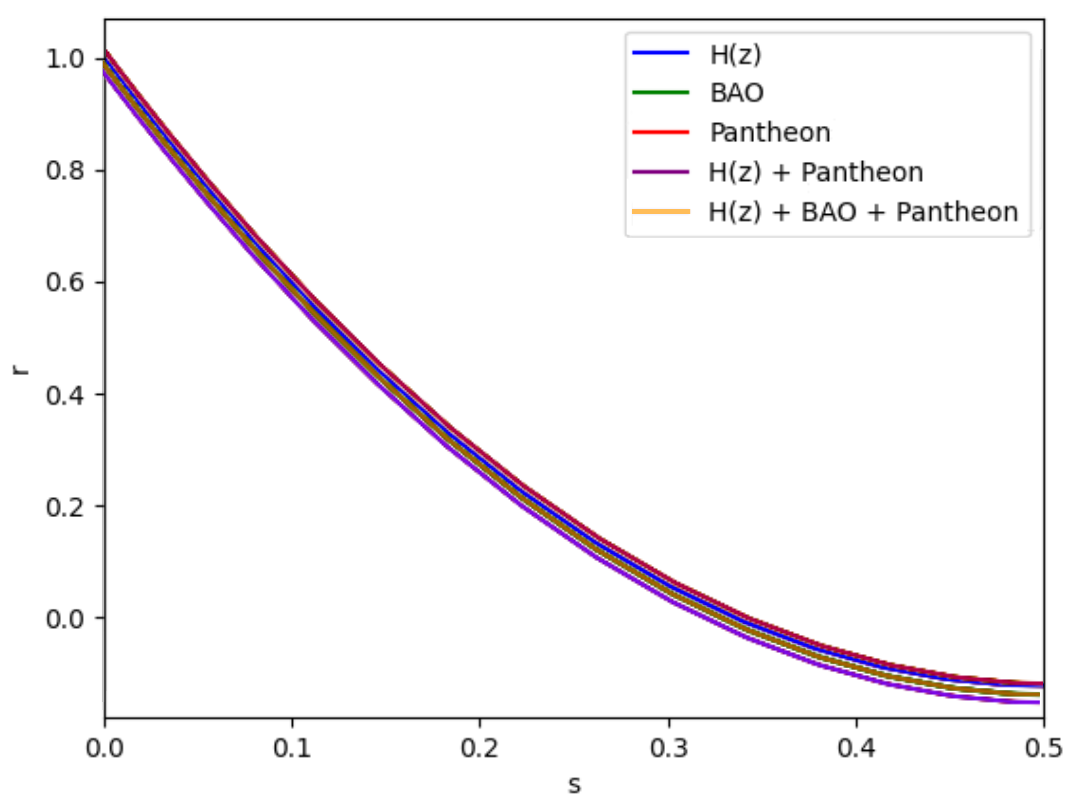}}
\caption{Figure depicts our model's $r$-$s$ plane behavior.}\label{r-s}
\end{figure}

Figure \ref{r-s} presents the evolutionary trajectory of the statefinder diagnostics $(r, s)$ for the proposed model. Based on Eqs. (\ref{eq28}) and (\ref{eq29}), the trajectory concludes at $(r, s) = (1, 0)$ when $z = -1$. This behavior aligns our model with the $\Lambda$CDM model during later evolutionary stages. The distinct paths of the $(r, s)$ parameters for the quintessence model, $\Lambda$CDM model, and CGDE model, as shown in Figure \ref{r-s}, provide evidence supporting the hypothesis of multiple phases of accelerated cosmic expansion.

\subsection{\textbf{The energy conditions}}
\noindent The physical validity of this model is determined by its energy conditions, which include the week energy condition (WEc), null energy condition (NEC), dominant energy condition (DEC), and strong energy condition (SEC). These conditions are fundamental to evaluating the feasibility of any cosmological model. By examining the current excess energy expenditure, we can assess the model's viability and gain insights into the nature of our universe. The energy conditions mentioned above are derived using Equations (\ref{eq10}) and (\ref{eq10p}).

\begin{equation}
WEC = 2^{(-1 + \gamma)} 3^{\gamma} \left(1 + (1 + z)^{2n} H_{0}^{2}\right)^{\gamma} \left(-1 + 12 \left(1 + (1 + z)^{2n} H_{0}^{2}\right)\right)
\end{equation}
\begin{equation}
    NEC = 2n (1 + z)^{2n} H_0^2 \left( 12 \left(1 + (1 + z)^{2n}\right) H_0^2 + 6^\gamma \left(\left(1 + (1 + z)^{2n}\right) H_0^2\right)^\gamma \right)
\end{equation}
\begin{equation}
    DEC =   24n \left( (1 + z)^{2n} + (1 + z)^{4n} \right) H_0^4 + 6^\gamma \left( \left(1 + (1 + z)^{2n} \right) H_0^2 \right)^\gamma \left( -1 + 2 \left( 6 - (-6 + n) (1 + z)^{2n} \right) H_0^2 \right)
    \end{equation}
\begin{equation}
SEC =   72n \left( (1 + z)^{2n} + (1 + z)^{4n} \right) H_0^4 + 6^\gamma \left( \left(1 + (1 + z)^{2n} \right) H_0^2 \right)^\gamma \left( 1 + 6 \left( -2 + (-2 + n) (1 + z)^{2n} \right) H_0^2 \right)
\end{equation}

\begin{figure}[H]
\centering
\includegraphics[scale=0.6]{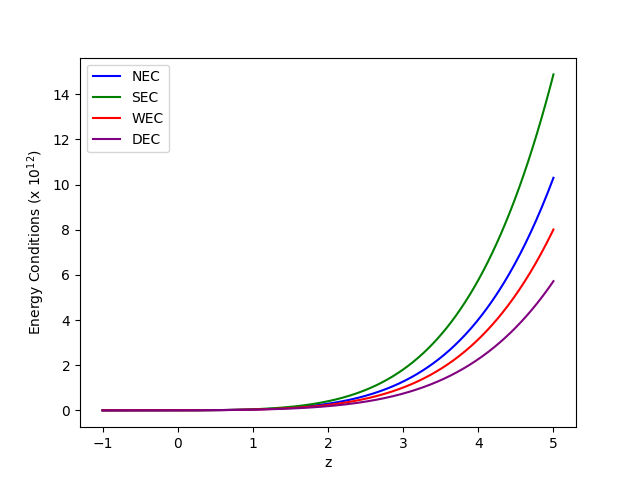}
\caption{This figure illustrates the enforcement of energy requirements for the $f(Q)$ gravity model and specifies that the y-axis values represent dimensionless quantities normalized by the critical density \( \rho_c \)}\label{ecs}
\end{figure}

\noindent The graphical behavior of the current cosmological model under WEC, NEC, DEC and SEC is exhibited in Fig. \ref{ecs}, considering an appropriate selection of constants. To ensure that gravity remains attractive and the gravitational field operates in a physically consistent manner, the Strong Energy Condition (SEC) must hold positive \cite{Yadav/2019bjp}.

\section{Concluding remarks}\label{V}
\noindent This study examines the dynamics of the universe within the framework of cubic $f(Q)$ gravity, using a simplified Hubble parameterization of the form $H = \frac{H_0}{\sqrt{2}} \left[1 + (1 + z)^{2n}\right]^{\frac{1}{2}}$. The free parameters of the model were constrained using observational data from Type Ia Supernovae (SN Ia). The results of the $\chi^2$ minimization procedure are shown in Table III. The analysis of the $H_0$ values reveals a $0.72\,\sigma$ tension when considering the OHD data alone, which reduces to $0.45\,\sigma$ when combining the OHD data with the Pantheon compilation of SN Ia. These results are in agreement with the findings of Planck \cite{Aghanim/2020}. Moreover, the aim of this research is to explore the accelerating expansion of the universe and to refine the estimation of cosmological parameters through the use of machine learning techniques. Artificial Neural Networks (ANN), Mixture Density Networks (MDN), and a new Multilayer Neural Network (MNN) were applied to improve the accuracy and efficiency of the parameter estimation process. By incorporating the Hubble parameter within the context of $f(Q)$ gravity, this approach helps address the $H_0$ tension. The study highlights several important findings:

\begin{table}
\caption{The model's parameters' estimated values (transposed).}
\renewcommand{\arraystretch}{1.8} 
\begin{center}
\begin{tabular}{|c|c|c|}
\hline
\textbf{Model/Dataset} & \boldmath$H_{0}$ \textbf{(km s\textsuperscript{-1} Mpc\textsuperscript{-1})} & \boldmath$n$ \\
\hline
OHD & $66.799^{+ 1.003}_{-0.897}$ & $1.298^{+ 0.112}_{- 0.092}$ \\
\hline
BAO & $66.692^{+ 0.817}_{-0.857}$ & $1.294^{+ 0.109}_{- 0.092}$ \\
\hline
Pantheon & $67.478^{+0.475}_{-0.485}$ & $1.296^{+ 0.111}_{- 0.110}$ \\
\hline
OHD + Pantheon & $66.429^{+0.495}_{-0.807}$ & $1.287^{+ 0.110}_{- 0.096}$ \\
\hline
OHD + BAO + Pantheon & $67.248^{+0.485}_{-0.817}$ & $1.297^{+ 0.108}_{- 0.107}$ \\
\hline
ANN & $67.054^{+ 0.443}_{-0.449}$ & $1.298^{+ 0.011}_{- 0.011}$ \\
\hline
MDN & $67.054^{+ 0.464}_{-0.464}$ & $1.298^{+ 0.011}_{- 0.011}$ \\
\hline
MNN & $67.033^{+ 0.439}_{-0.464}$ & $1.298^{+ 0.011}_{- 0.011}$ \\
\hline
\end{tabular}
\end{center}
\end{table}
\begin{itemize}
\item[i)] The energy density of the universe under cubic $f(Q)$ gravity evolves with positive sign and the cosmic pressure and energy density decrease as $z \rightarrow 0$.  Fig. \ref{gamma3} displays the energy density $\rho$ and cosmic pressure $p$ of the model.

\item[ii)] The EoS parameter $\omega$ evolves with positive sign for $z \geq 0.59$ and it flips to negative sign when $z < 0.59$. According to Fig \ref{eos}, the model's EoS parameter is now in the quintessence period and indicates a future affinity for the $\Lambda$CDM world.

\item[iii)] CoLFI successfully computed cosmological parameters by employing Artificial Neural Networks (ANNs), Mixture Density Networks (MDNs), and Multilayer Neural Networks (MNNs). By learning the conditional probability densities from observational data and posterior distributions, the process of parameter estimation became more efficient. The accuracy of the parameters and the training efficiency of the neural networks were improved through the use of hyperellipsoid parameters. In terms of performance, MNN demonstrated comparable results to Markov Chain Monte Carlo (MCMC) methods, confirming its effectiveness and reliability.

\item[iv)] In the context of the model, the energy density decreases as the redshift approaches zero, maintaining a positive value throughout the evolution of the universe. Graphs depict the temporal variation of $\rho$. As redshift decreases, pressure similarly diminishes, reflecting the ongoing expansion and acceleration of the universe. A negative $\omega$ value at $z \geq 0.59$ signals the transition of the universe from deceleration to acceleration at lower redshifts. The confirmation of the universe's late-time acceleration is evident when the deceleration parameter $q$ crosses zero, marking a phase transition from deceleration to acceleration.

\item[v)] A direct comparison between our model and the standard \( \Lambda \)CDM framework has been illustrated in Fig. \ref{FR1} through the evolution of \( H(z) \) and \( \mu(z) \), using both MCMC and deep learning approaches. While the parameter \( \gamma \) modulates the form of \( f(Q) \), the observational constraints and overall cosmological behavior remain anchored to the same Hubble parameterization and data sets. Therefore, the comparative trends observed in Figure 4 continue to hold qualitatively across different values of \( \gamma \), offering meaningful insight into the model's compatibility with \( \Lambda \)CDM. 

\end{itemize}

\noindent The proposed cosmos model facilitates late universe acceleration in the absence of dark energy and circumvents issues associated with the cosmological constant. Based on the observable data, a thorough examination of modeling a $f(Q)$ gravity is necessary. Neural network methods were identified as viable alternatives for estimating cosmological parameters, replacing the traditional MCMC approach. The $f(Q)$ gravity model integrates ANN, MDN, and MNN methodologies, facilitating a comprehensive analysis of competing gravity theories and their implications for the expansion of the universe. This study is an integral component of an investigation into the intersection of machine learning and cosmology. This highlights the potential of machine learning in tackling complex cosmic dynamics and expansion challenges. The examination of density, cosmic pressure, the equation of state, and deceleration leads to robust conclusions.

\section*{Acknowledgement}
\noindent The authors are very grateful to the honorable referee and the editor for illuminating suggestions that have significantly improved our work in terms of research quality and presentation.

\newpage
\appendix
\section{}
    \begin{figure}[htbp]
        \centering
        \includegraphics[width=0.9\textwidth]{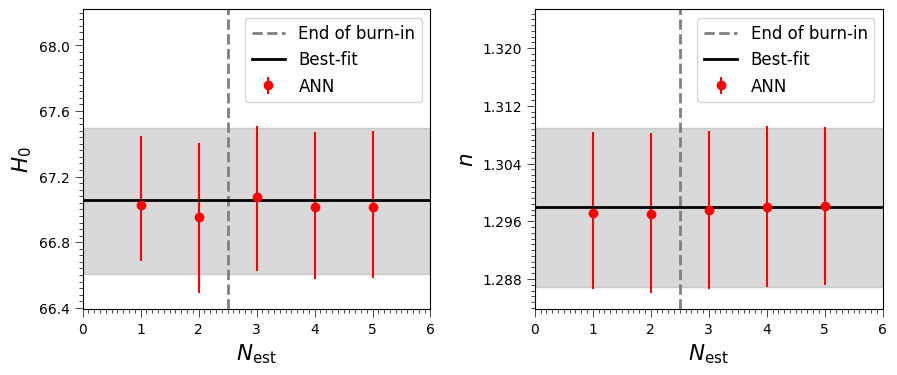}
        \caption{The values that are the best match for the cosmological parameters and the errors of one standard deviation are provided as a function of the steps. The results from the ANN method are shown as red circles with error bars; solid black lines and a grey-shaded area depicts the best-fit values}
        \label{fig:sub1}
    \end{figure}
    \begin{figure}[htbp]        
    \centering
        \includegraphics[width = \textwidth]{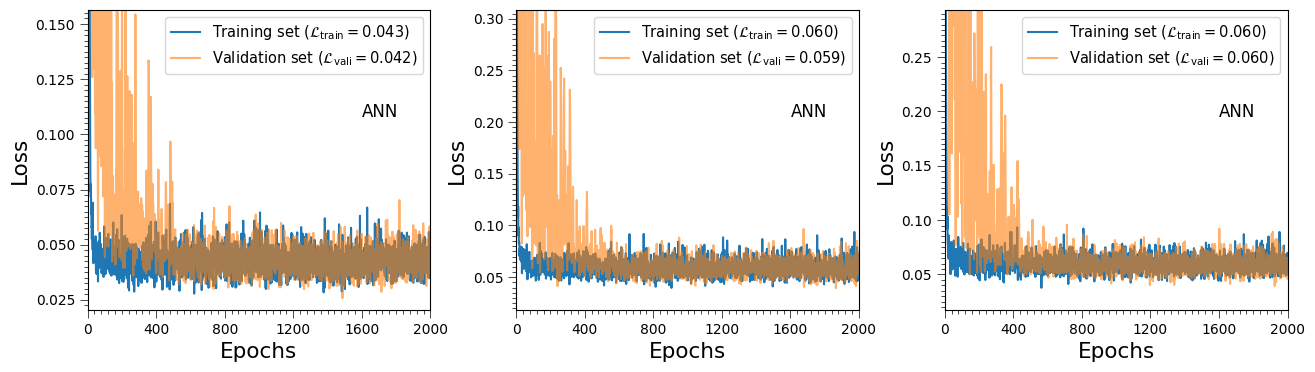}
        \caption{Losses of the training set and validation set are shown. The training set consists of 3000 samples, while the validation set comprises 500 samples.}
        \label{fig:sub3}
    \end{figure}

\begin{figure}[htbp]
    \centering
        \centering
        \includegraphics[width=0.7\textwidth]{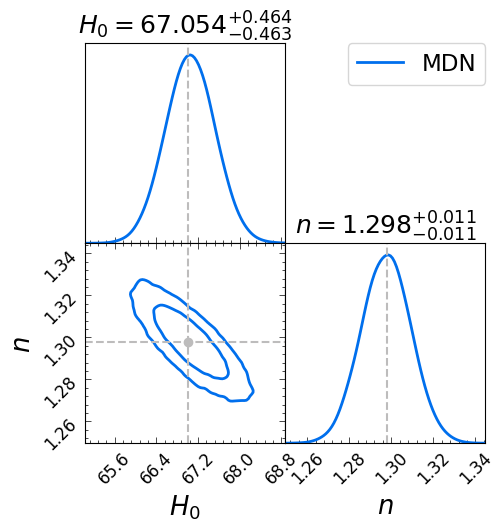}
        \caption{One-dimensional and two-dimensional marginalized distributions of $H_0$ and $n$, showing the 1$\sigma$ and 2$\sigma$ contours. These constraints are derived from 57-point $H(z)$ data, trained and simulated using the MDN model.}
        \label{fig:sub2}
    \end{figure}
    \begin{figure}[htbp]
        \centering
        \includegraphics[width=0.7\textwidth]{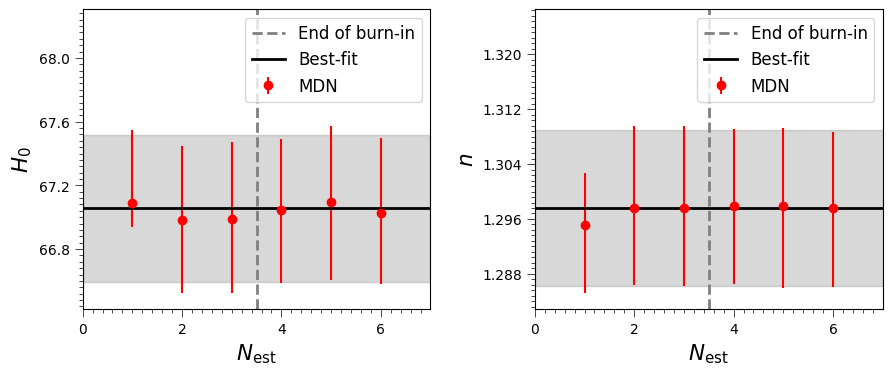}
        \caption{The best-fit values and 1$\sigma$ errors of cosmological parameters as a function of steps are presented. The results from the MDN method are shown as red circles with error bars; solid black lines and grey-shaded areas depict the best-fit values}
        \label{fig:sub1}
    \end{figure}
    \begin{figure}
        \centering
        \includegraphics[width = \textwidth]{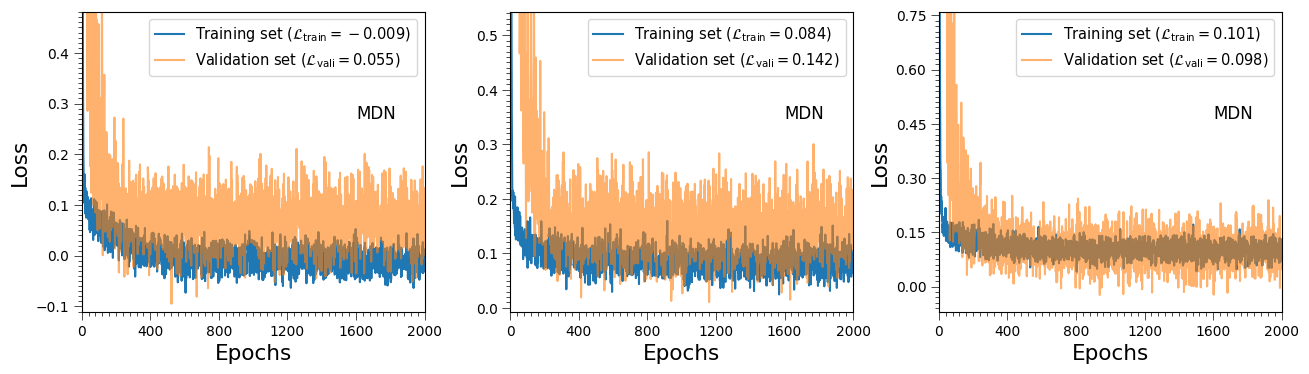}
        \caption{Losses of the training set and validation set are shown. The training set consists of 3000 samples, while the validation set comprises 500 samples.}
        \label{fig:sub3}
    \end{figure}
    
\begin{figure}[htbp]
\centering
\includegraphics[width=0.7\textwidth]{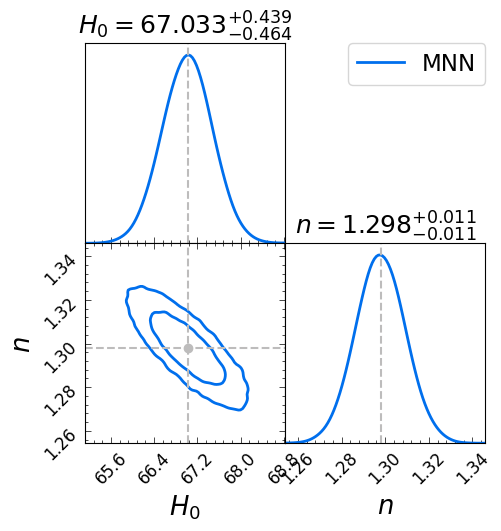}
\caption{One-dimensional and two-dimensional marginalized distributions of $H_0$ and $n$, showing the 1$\sigma$ and 2$\sigma$ contours. These constraints are derived from 57-point $H(z)$ data, trained and simulated using the MNN model.
}
\label{fig:sub2}
 \end{figure}
    
    \begin{figure}[htbp]
        \centering
        \includegraphics[width=0.7\textwidth]{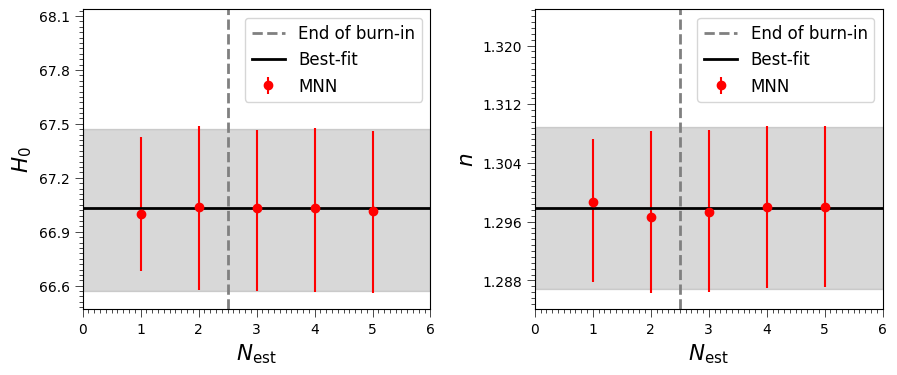}
        \caption{The best-fit values and 1$\sigma$ errors of cosmological parameters as a function of steps are presented. The results from the MNN method are shown as red circles with error bars; solid black lines and grey-shaded areas depict the best-fit values}
        \label{fig:sub1}
    \end{figure}
    
    \begin{figure}[htbp]
        \centering
        \includegraphics[width = \textwidth]{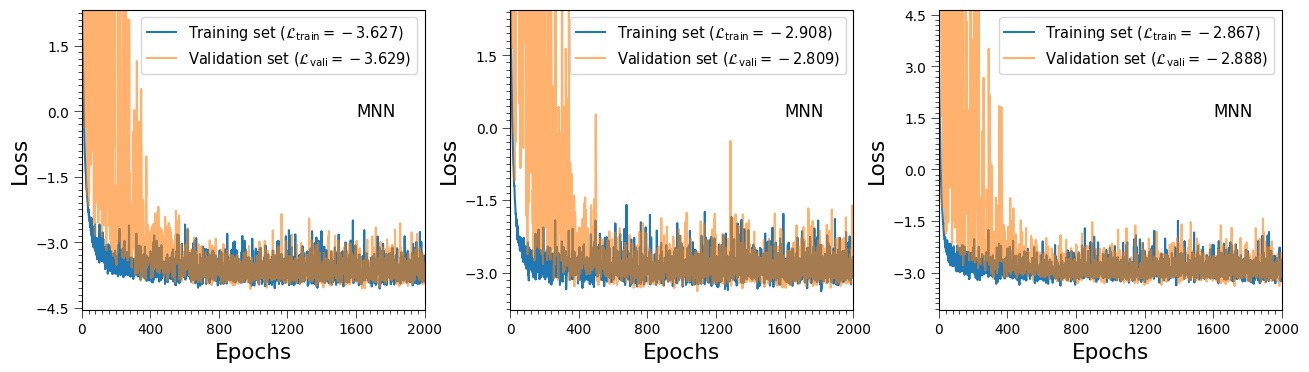}
        \caption{Losses of the training set and validation set are shown. The training set consists of 3000 samples, while the validation set comprises 500 samples.}
        \label{fig:sub3}
    \end{figure}


\end{document}